\documentclass[acmsmall]{acmart}
\newtheorem{asm}{Assumption}

\newtheorem{rem}{Remark}
\newtheorem{obj}{Safe Objective}

\begin{document}

\title{S$\mathcal{L}_{1}$-Simplex: Safe Velocity Regulation of Self-Driving Vehicles in Dynamic and Unforeseen Environments}

\author{Yanbing Mao}
\affiliation{%
  \institution{University of Illinois at Urbana--Champaign}
  \streetaddress{Department of Mechanical Science and Engineering}
  \country{USA}}
\email{ybmao@illinois.edu}

\author{Yuliang Gu}
\affiliation{%
  \institution{University of Illinois at Urbana--Champaign}
  \streetaddress{Department of Mechanical Science and Engineering}
  \country{USA}}
\email{yuliang3@illinois.edu}

\author{Naira Hovakimyan}
\affiliation{%
  \institution{University of Illinois at Urbana--Champaign}
  \streetaddress{Department of Mechanical Science and Engineering}
  \country{USA}}
\email{nhovakim@illinois.edu}

\author{Lui Sha}
\affiliation{%
  \institution{University of Illinois at Urbana--Champaign}
  \streetaddress{Department of Computer Science}
  \country{USA}}
\email{lrs@illinois.edu}

\author{Petros Voulgaris}
\affiliation{%
  \institution{University of Nevada}
  \streetaddress{Department of Mechanical Engineering}
 \country{USA}}
\email{pvoulgaris@unr.edu}

\renewcommand{\shortauthors}{Mao and Gu, et al.}

\begin{abstract}
This paper proposes a novel extension of the Simplex architecture with model switching and model learning to achieve safe velocity regulation of self-driving vehicles in dynamic and unforeseen environments. To guarantee the reliability of autonomous vehicles, an $\mathcal{L}_{1}$ adaptive controller that compensates for uncertainties and disturbances is employed by the Simplex architecture as a verified high-assurance controller (HAC) to tolerate concurrent software and physical failures. Meanwhile, the safe switching controller is incorporated into the HAC for safe velocity regulation in the dynamic (prepared) environments, through the integration of the traction control system and anti-lock braking system. Due to the high dependence of vehicle dynamics on the driving environments, the HAC leverages the finite-time model learning to timely learn and update the vehicle model for $\mathcal{L}_{1}$ adaptive controller, when any deviation from the safety envelope or the uncertainty measurement threshold occurs in the unforeseen driving environments. With the integration of $\mathcal{L}_{1}$ adaptive controller, safe switching controller and finite-time model learning, the vehicle's angular and longitudinal velocities can asymptotically track the provided references in the dynamic and unforeseen driving environments, while the wheel slips are restricted to safety envelopes to prevent slipping and sliding. Finally, the effectiveness of the proposed Simplex architecture for safe velocity regulation is validated by the AutoRally platform. Demonstration video available at: \textcolor[rgb]{1.00,0.00,1.00}{\url{https://ymao578.github.io/pubs/m1.mp4}}
\end{abstract}


\maketitle

\section{Introduction}
Intelligent transportation systems (ITS) that embed vehicles, roads, traffic lights, message signs, along with microchips and sensors are bringing significant improvements in transportation system performance, including reduced congestion, increased safety and traveler convenience \cite{dimitrakopoulos2010intelligent}. Intelligent vehicles that aim to improve traffic safety, transport efficiency and driving comfort are playing a major role in the ITS, among which the
longitudinal vehicle dynamics control is an important aspect. Traction control system (TCS) and anti-lock braking system (ABS) are representative technologies for longitudinal vehicle dynamic control systems \cite{ivanov2014survey}. Specifically, the ABS is primarily designed to prevent excessive or insufficient wheel slip to keep vehicle steerable and stable during intense braking events, which contributes to high brake performance and road safety \cite{savaresi2010active, aly2011antilock,reif2014brakes}, while the TCS primarily regulates wheel slip to reduce or eliminate excessive slipping or sliding during vehicle acceleration, which results in better drivability, safety and traction performance in adverse weather or traffic conditions \cite{borrelli2006mpc,de1999dynamic,colli2006single,reichensdorfer2018stability,yin2009novel}. Both TCS and ABS are complicated by nonlinearities, uncertainties and parameter variations, which are induced by variations in the disc pad friction coefficient \cite{han2017accurate}, nonlinear relation between brake torque and pressure \cite{han2017accurate}, nonlinear wheel-slip characteristics \cite{de1999dynamic}, among many others. To regulate slip in these challenging scenarios, various model-based control schemes have been proposed, e.g., proportional-integral-derivative control  in combination with sliding mode observer \cite{magallan2010maximization}, fuzzy control \cite{khatun2003application}, model predictive control \cite{borrelli2006mpc}, sliding mode control \cite{han2017development}, $H_{\infty}$ control \cite{de2017comparison}.

Autonomous velocity regulation has gained a vital importance \cite{dias2014longitudinal, sun2019optimal,tai2000robust}, which is motivated by, for example, the imposed speed limits on driving zones (see e.g., school zone and commercial street) and required relative positions with respect to surrounding vehicles and obstacles for safety and transport efficiency. Velocity regulation needs the vehicle to operate in either drive or brake mode. However, the current frameworks ignore the wheel slip regulation for safety \cite{dias2014longitudinal, sun2019optimal,tai2000robust}, which has always been a common control objective of TCS and ABS. Therefore, the concurrent velocity and slip regulations
are indispensable for enhanced safety, drivability, stability and steerability \cite{savaresi2010active, aly2011antilock,reif2014brakes,borrelli2006mpc,de1999dynamic,colli2006single,reichensdorfer2018stability,yin2009novel}. Inspired by these observations, this paper focuses on \emph{safe velocity regulation} through integrating the TCS and ABS. More concretely, the vehicle asymptotically steers its angular and longitudinal velocities to the provided references, while restricts its wheel slips to the safety envelopes to prevent slipping and sliding during intense braking and accelerating events.

However, as a typical cyber-physical system, self-driving vehicles integrate the vehicular cyber system with the vehicular physical system and the environmental model for control and operation, whose increasing complexity hinders its reliability, especially when system failures occur. The Simplex architecture -- using simplicity to control complexity -- provides a reliable control system via software approach, whose core idea is to tolerate control software failures \cite{sha2001using}. For a self-driving vehicle, its complicated control missions (e.g., traction control and parallel parking) exacerbate the difficulty of keeping the system safe in the presence of physical failures, since the control actuation computed in cyber layer depends on the physical modeling. Moreover, the inaccurate vehicle and/or tire parameters are the main obstacles for preventing wheel slip-based control in TCS and ABS. Hence, the Simplex architecture needs an adaptive controller -- compensating for model and parameter uncertainties -- as a verified safe controller to tolerate physical failures as well. Among the various adaptive control methods, $\mathcal{L}_{1}$ adaptive controller has been widely adopted due to its fast adaptation, guaranteed robustness, and predictable transient response \cite{hovakimyan2010LL1,hovakimyan201154}. Considering that $\mathcal{L}_{1}$ adaptive control has been verified consistently with the theory in dealing with physical failures with transient performance and robustness guarantees \cite{ackerman2016l1,leman2009l1,choe2011handling}, Wang et al. in  \cite{wang2018rsimplex} proposed an  $\mathcal{L}_{1}$-Simplex to tolerate concurrent software and physical failures. Inspired by the attractive properties of $\mathcal{L}_{1}$-Simplex, this paper proposes a variant of $\mathcal{L}_{1}$-Simplex for \emph{safe velocity regulation} of self-driving vehicles, where $\mathcal{L}_{1}$ adaptive controller works as a verified safe controller that compensates for uncertainties, disturbances, software and physical failures.

One of the fundamental assumptions of model-based controllers is the availability of a relatively accurate model of the underlying dynamics in consideration. However, vehicle dynamics highly depend on the driving environments \cite{borrelli2006mpc,de1999dynamic}, and can be significantly different from one road (e.g., asphalt) to another one (e.g., snow). Therefore, a single off-line-built vehicle model cannot capture the  differences in the dynamics induced by environmental variations. To address the model mismatch issue, we bring switching control scheme into $\mathcal{L}_{1}$-Simplex, where multiple off-line-built models that correspond to different environments  (e.g., snow and icy) are stored in $\mathcal{L}_{1}$ adaptive control architecture, thus yielding the switching $\mathcal{L}_{1}$ adaptive controller. The switching $\mathcal{L}_{1}$ adaptive controller aims at the safe velocity regulation in the dynamically changing environmental conditions that can be modeled offline, where each model's remaining mismatch can be compensated by the $\mathcal{L}_{1}$ adaptive controller.

Due to the high dependence of vehicle dynamics on driving environments (including, e.g., air mass density, wind velocity and road friction coefficient \cite{rajamani2011vehicle}), it is unreasonable to expect that the off-line-built multiple models are sufficient to accurately describe the vehicle-environment interaction dynamics in an unforeseen or unprepared environment, as e.g. the 2019 New York City Snow Squall \cite{software}. When the unforeseen environments cause deviation from the safety envelope or the uncertainty measurement threshold in the time-critical environment, timely learning and updating the vehicle model using most recent sensor data (generated in the unforeseen environment) is indispensable for safe velocity regulation. To address the problem, we incorporate finite-time model learning into $\mathcal{L}_{1}$-Simplex, which can timely learn and update a vehicle model for $\mathcal{L}_{1}$ adaptive controller in the unforeseen driving environments.

To this end, we propose a novel Switching $\mathcal{L}_{1}$-Simplex architecture (S$\mathcal{L}_{1}$-Simplex) \textcolor[rgb]{0.00,0.00,1.00}{with the novel incorporation of switching $\mathcal{L}_{1}$ adaptive controller and finite-time model learning for self-driving vehicles}, which is able to achieve
\begin{itemize}
  \item \textcolor[rgb]{0.00,0.00,1.00}{safe velocity regulation in the dynamic and unforeseen driving environments,}
  \item \textcolor[rgb]{0.00,0.00,1.00}{safety envelop extending,}
  \item tolerance of concurrent software and physical failures.
\end{itemize}

This paper is organized as follows. In Section 2, we present the preliminaries including longitudinal vehicle model and the S$\mathcal{L}_{1}$-Simplex architecture. The safety envelope is formulated in Section 3.  In Section 4, we present the off-line-built vehicle models and the finite-time model learning procedure, based on which, we present the S$\mathcal{L}_{1}$-Simplex design in Section 5. We present the experiments in Section 6. We finally present our conclusions and future research directions in Section 7.

\section{PRELIMINARIES}
\subsection{Notation}
We let $\mathbb{R}^{2}$ denote the set of two dimensional real vectors. $\mathbb{R}^{2 \times 2}$ denotes the set of $2 \times 2$-dimensional real matrices. $\mathbb{N}$ stands for the set of natural numbers, and $\mathbb{N}_{0} = \mathbb{N} \cup \{0\}$. ${\mathbb{V}} \setminus \mathbb{K}$ stands for the complement set of $\mathbb{K}$ with respect to $\mathbb{V}$. $\mathbf{I}$ and $\mathbf{1}$, respectively,  denote the identity matrix and the vector of all ones, with proper dimensions. For $x \in \mathbb{R}^{2}$, $\left\| x \right\| = \sqrt {x_1^2 + x_2^2}$. For $A \in \mathbb{R}^{2 \times 2}$, $\left\| A \right\|$ denotes the induced 2-norm of a matrix $A$,   $|| A ||_{\mathrm{F}}$ denotes the Frobenius norm of matrix $A$. The superscript `$\top$' stands for matrix transpose.  $|\mathbb{T}|$ denotes the cardinality (i.e., size) of set $\mathbb{T}$. We use $P$ $>$ $(<)$ 0 to denote a positive definite (negative definite) matrix $P$. Given a symmetric matrix $P$, $\lambda_{\min}(P)$ and $\lambda_{\max}(P)$ are the minimum and maximum eigenvalues, respectively. $\mathcal{L}_{1}$ norm of a function $x(t)$ is denoted by $\left\| x(t) \right\|_{\mathcal{L}_{1}}$, and ${\left\| x \right\|_{{\mathcal{L}_\infty }\left[ {a,b} \right]}} = {\sup _{a \le t \le b}}\left\| {x( t)} \right\|$. We denote $x( s ) = \mathfrak{L}\left\{x ( t ) \right\}$, where $\mathfrak{L}(\cdot)$ stands for the Laplace transform operator. The gradient of $f(x)$ at $x$ is denoted by $\nabla f( x )$.

\subsection{Switching $\mathcal{L}_{1}$-Simplex Architecture}
In this subsection, we introduce the Simplex architecture with incorporation of safe switching control and finite-time model learning, which is adopted
from $\mathcal{L}_{1}$-Simplex proposed in \cite{wang2018rsimplex}. We first present the assumption on the Simplex architecture for self-driving vehicles.
\begin{asm}
The vehicle is equipped with sensors for real-time environmental perception, which can accurately detect the driving environments. \label{asm2}
\end{asm}

\begin{figure}[http]
\centering{
\includegraphics[scale=0.33]{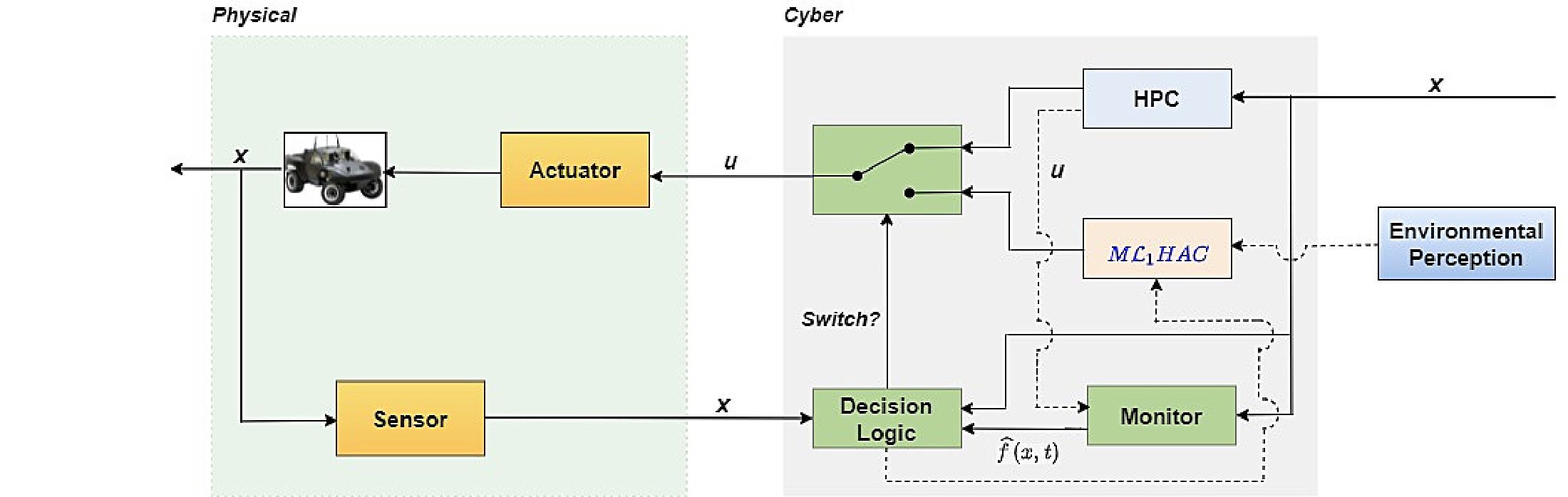}
}
\caption{S$\mathcal{L}_{1}$-Simplex architecture.}
\label{RSimplex}
\end{figure}
\begin{figure}[http]
\centering{
\includegraphics[scale=0.44]{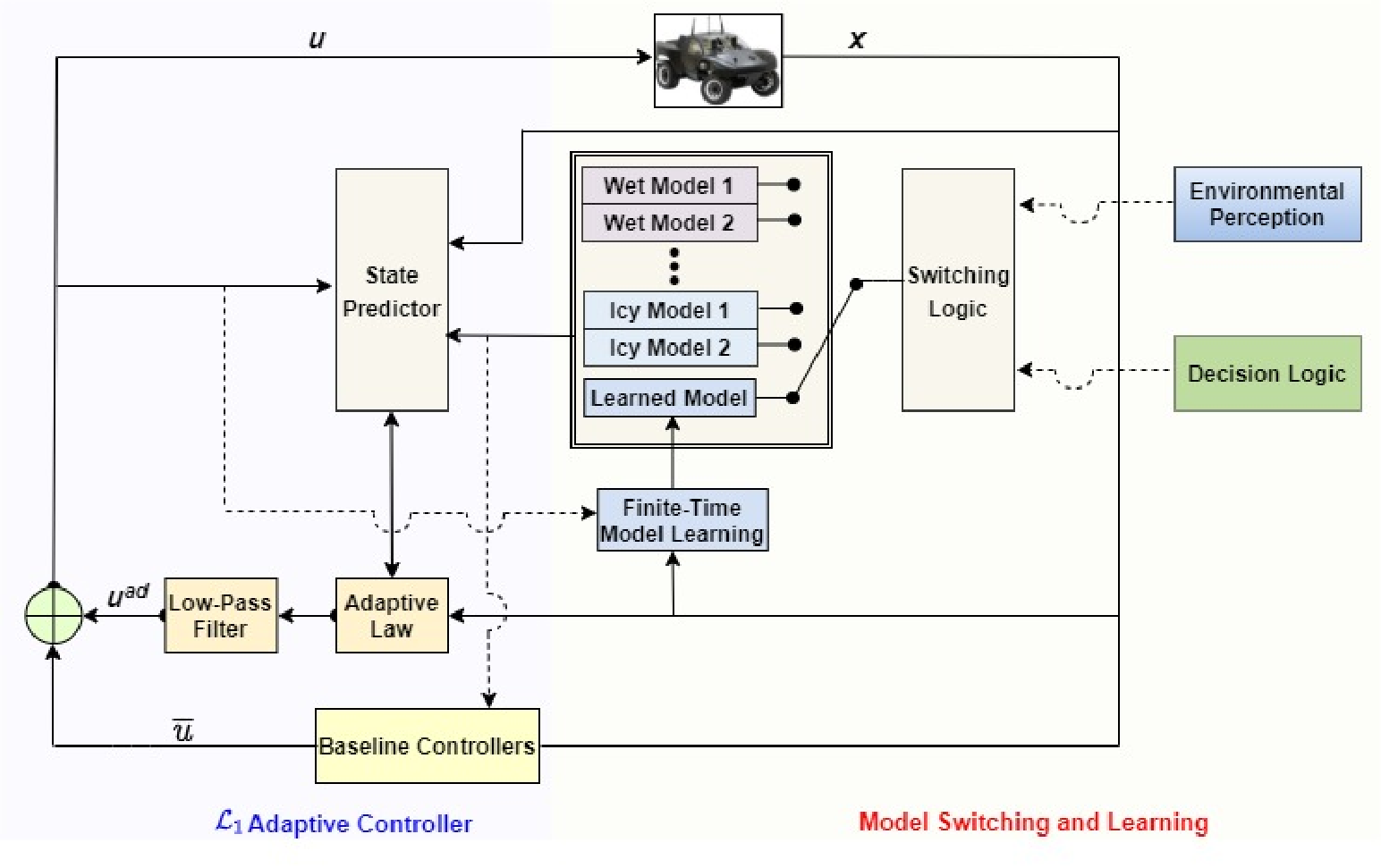}
}
\caption{M$\mathcal{L}_{1}$HAC: $\mathcal{L}_{1}$-based HAC architecture with switching control and finite-time model learning.}
\label{RHAC}
\end{figure}

As described by Fig. \ref{RSimplex}, the proposed S$\mathcal{L}_{1}$-Simplex architecture for self-driving vehicles includes
\begin{itemize}
\item High-Performance Controller (HPC): The HPC is a complex controller, which provides high levels of performance and advanced functionalities (e.g., the cautious model predictive control \cite{hewing2019cautious}, $\mathcal{L}_1-\mathcal{GP}$ \cite{gahlawat2020l1}  and the end-to-end control via variational autoencoder \cite{amini2018variational}), and is active during normal operation of the system and possibly not fully verified.
\item Model Learning and $\mathcal{L}_{1}$ Based High-Assurance Controller (M$\mathcal{L}_{1}$HAC): The M$\mathcal{L}_{1}$HAC is a \emph{simple and verified} controller that provides limited levels of performance and reduced functionalities to guarantee safe and stable operation of the vehicle. As shown in Fig. \ref{RHAC}, the M$\mathcal{L}_{1}$HAC includes
  \begin{itemize}
   \item the $\mathcal{L}_{1}$ adaptive controller, which compensates for uncertainties, disturbances, software and physical failures for velocity regulation;
    \item the stored off-line-built vehicle models (obtained via, e.g., Gaussian process regression \cite{hewing2019cautious}) that vary with environments, which guarantee safe velocity regulation in the dynamic normal (known and prepared) driving environments;
    \item the finite-time model learning, which timely learns and updates the vehicle model for safe velocity regulation in the unforeseen driving environments;
    \item the switching logic that depends on the environmental perception and the real-time verification of safety envelope, which is responsible for activating an off-line-built model or on-line learned model for $\mathcal{L}_{1}$ adaptive controller.
  \end{itemize}
  \item Uncertainty Monitor: This \emph{verified} monitor takes the form of the state predictor in $\mathcal{L}_{1}$ adaptive control architecture, which provides estimates of the uncertainties inside the vehicle system with fast adaptation.
  \item Decision Logic: This \emph{verified} logic depends on the magnitudes of uncertainty estimations and the real-time verification of the safety envelope, which triggers the switching from HPC to M$\mathcal{L}_{1}$HAC in the events of software and/or physical failures and/or large model mismatch occurrence.
\end{itemize}

\begin{rem}
In the proposed Simplex architecture, finite-time model learning is running in parallel with M$\mathcal{L}_{1}$HAC and HPC, which is depicted in Fig. \ref{RHAC}. This configuration guarantees that when model learning is needed for reliable decision making, a model that corresponds to the current operating environment is available immediately, so that M$\mathcal{L}_{1}$HAC is always in control. If operating without the configuration of parallel running, the car can lose the control in the unforeseen environments due to the time delay in collecting state samplings and learning.
\end{rem}

As shown in Figs. \ref{RSimplex} and \ref{RHAC},  the proposed Simplex includes three types of switching: 1) switching between HPC and M$\mathcal{L}_{1}$HAC, 2) switching between stored vehicle models and learned vehicle models, and 3) switching between two subsystems in a fixed normal environment. Therefore, excluding Zeno behaviors is needed to guarantee the feasibility of the proposed framework. To achieve this, we impose a minimum dwell time ${\mathrm{dwell}_{\min }}$ on HPC, M$\mathcal{L}_{1}$HAC, learned vehicle models and stored sub-models, i.e.,
\begin{align}
\mathop {\min }\limits_{\forall k \in {\mathbb{N}_0}} \left\{ {{t_{k + 1}} - {t_k}} \right\} \ge {\mathrm{dwell}_{\min }} > 0,\label{minimumdwelltime}
\end{align}
where $t_{k}$ denotes a switching time.

\subsection{Safe Objectives}
The proposed Simplex has two objectives of safe control, which are formally stated below.
\begin{obj}
The vehicle asymptotically steers its angular and longitudinal velocities to the provided references, while restricts its wheel slips to the safety envelopes in the dynamic and unforeseen driving environments.\label{obj1}
\end{obj}
\begin{obj}
The vehicle control system tolerates the concurrent software and physical failures.\label{obj2}
\end{obj}

\subsection{Vehicle Model}
\begin{figure}[http]
\centering{
\includegraphics[scale=0.45]{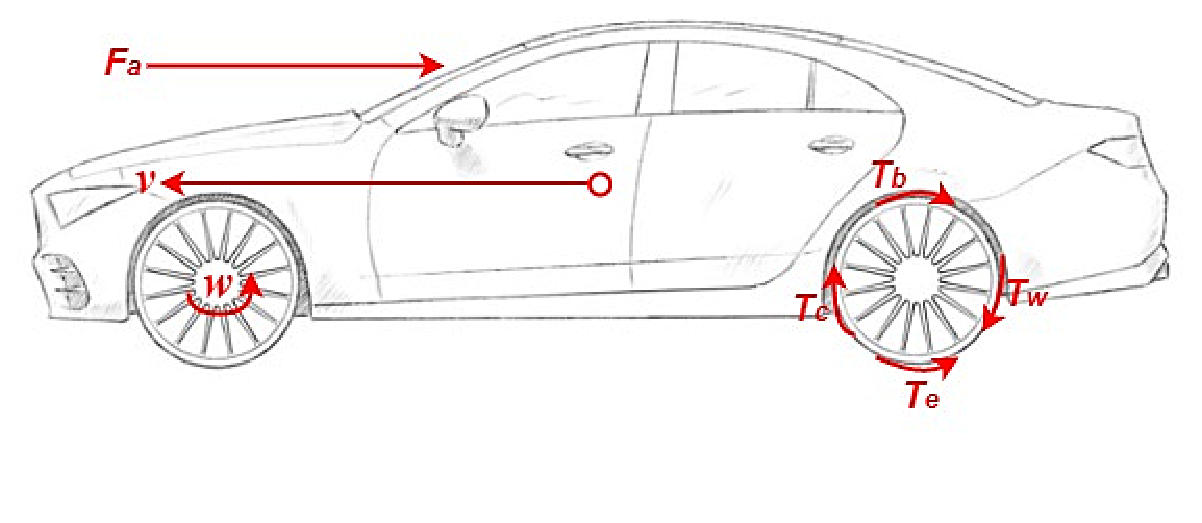}
}
\caption{Front-wheel-driven vehicle model.}
\label{vehiclemodel}
\end{figure}

Moving forward, we present the following assumption pertaining to vehicle longitudinal model for our model-based control systems: TCS and ABS. The parameter notations of vehicle model are given in Table 1.
\begin{table}[ht]
\centering
\caption{Vehicle Model Parameters}
\begin{tabular}{|l|l|}
\hline
$w$  &   Angular velocity         \\ \hline
$v$   &  Longitudinal velocity       \\ \hline
$J$ &   Wheel rotational inertia  \\ \hline
$T_{b}$  & Brake torque         \\ \hline
$T_{c}$  & Friction torque on wheel   \\ \hline
$T_{e}$  & Engine torque  \\ \hline
$T_{w}$  & Viscous torque on wheel   \\ \hline
$F_{a}$ &   Longitudinal aerodynamic drag force \\ \hline
$\zeta$ & Aerodynamic drag constant \\ \hline
$\varrho$ & Viscous friction in driven wheel\\ \hline
$P$  & Master cylinder pressure   \\ \hline
$r$ & Wheel radius \\ \hline
$m$ & Vehicle mass \\ \hline
$C$ &  Brake piston effective area \\ \hline
$\eta _b$ &  Pad friction coefficient \\ \hline
${r_b}$ &  Brake disc effective radii \\ \hline
$h$ &   Gravity center height  \\ \hline
\end{tabular}
\end{table}

\begin{asm}The vehicle's
\begin{itemize}
  \item  dynamics of the left and the right sides are identical (i.e. the vehicle is symmetric);
  \item  wheel is damped with a viscous torque \cite{kirchner2011anthropomimetic}, i.e.,
  \begin{align}
  T_{w}(t) =  \varrho w(t); \label{viscoustorque}
  \end{align}
  \item  longitudinal aerodynamic drag force can be linearized in term of longitudinal velocity \cite{kirchner2011anthropomimetic}, i.e.,
  \begin{align}
  F_{a}(t) =  \zeta v(t). \label{aerodynamicforce}
  \end{align}
\end{itemize}\label{asm1}
\end{asm}

The longitudinal vehicle model is depicted in Fig. \ref{vehiclemodel},  whose control variables are engine torque and master cylinder pressure. The uncertainties pertaining to the relations \eqref{viscoustorque} and \eqref{aerodynamicforce}  are included in the following dynamics of the vehicle's longitudinal and wheel motions on a flat road, \cite{rajamani2011vehicle}:
\begin{subequations}
\begin{align}
J\dot w(t) &= {T_e}(t) - {T_w}(t) - {T_b}(t) - {T_c}(t) + \tilde{f}_{w}(t), \label{carmodel1}\\
m\dot v(t) &= \frac{{{T_c}(t)}}{r} - {F_{a}}(t) + \tilde{f}_{v}(t), \label{carmodel2}
\end{align}\label{carmodel}
\end{subequations}
\!\!\!where $\tilde{f}_{w}(t)$ and $\tilde{f}_{v}(t)$ represent uncertainties that are due to the modeling errors, noise, disturbances, unmodeled forces/torques in \eqref{viscoustorque}, \eqref{vehiclemodel}, and others.

Following \cite{han2017accurate}, the actual relation between the master cylinder pressure and the brake torque is modeled by a linear model with uncertainty:
\begin{align}
{T_b}(t) = C{\eta _b}{r_b}P(t) + \varpi_{b}( t ), \label{pressure}
\end{align}
where the unknown $\varpi_{b}( t )$ denotes the uncertainty.

The wheel slip is defined in term of wheel and longitudinal velocities as
\begin{align}
s(t) = \left| {v(t) - rw(t)} \right|. \label{slipdef}
\end{align}
We note that $s(t) = 0$ and $s(t) = \max \left\{ {v(t),rw(t)} \right\}$ indicate pure rolling and full sliding, respectively. In this paper, the slip will be imposed on velocity tracking control as a safety constraint.

We now define a set of environmental model indices:
\begin{align}
\mathbb{E} = \{\mathrm{dry}, ~\mathrm{wet}, ~\mathrm{snow}, ~\ldots, ~\mathrm{icy}, ~\mathrm{learned}_{1}, ~\mathrm{learned}_{2}, ~ \ldots,~\mathrm{learned}_{q}\},\label{environmentset}
\end{align}
for which, we further define a subset:
\begin{align}
\mathbb{L} = \{\mathrm{learned}_{1}, ~\mathrm{learned}_{2}, ~\ldots,~\mathrm{learned}_{q}\},\label{learnset}
\end{align}
which denotes a set of learned models an unforeseen driving environment, wherein the $\mathcal{L}_{1}$ controller relies on the learned models. The $\mathbb{L}$ can also be used to indicate an unforeseen driving environment in the paper.

\begin{rem}
\textcolor[rgb]{0.00,0.00,1.00}{In our proposed framework, the learned model is not necessarily a single one. For example, assuming the learned model comes from the data generated in the slipping mode, when working in the skidding mode, if the (learned) model mismatch therein leads to a deviation from the safety envelope or the uncertainty measurement threshold, the finite-time model learning will be triggered to immediately output a learned model to replace the previous leaned one in M$\mathcal{L}_{1}$HAC. In the most ideal scenario (as the one in our experiment section), if the model mismatch never triggers the model learning, the Simplex only uses the first learned model. The ideal scenario means the first learned model can capture critical system properties in both slipping and skidding modes, and the $\mathcal{L}_{1}$ adaptive controller then compensates for the un-captured system properties, such that the single model is sufficient to enhance safety assurance in the ideal scenarios.}
\end{rem}

The experimental data of tire friction models shows that the tire friction torque depends on slip (or slip ratio) and road friction coefficient \cite{de1999dynamic, borrelli2006mpc}. We also use a linear model with uncertainty to describe the actual relation between the tire friction torque, slip and road friction coefficient, i.e., ${T_c}( t) = {k_{\sigma ( t )}}s(t) + \varpi_{c}( t )$,  where
${k_{\sigma ( t )}}$ is obtained from experimental data via parameter identification, $\varpi_{c}( t )$ denotes the unknown uncertainty, and $\sigma(t) \in \mathbb{E}$, where, e.g., $\sigma(t)$ = `\text{snow}` for $t \in [t_{k}, t_{k+1})$ means the vehicle is driving in the snow environment during the time interval $[t_{k}, t_{k+1})$. With the consideration of \eqref{slipdef}, ${T_c}( t)$ is equivalently expressed as
\begin{align}
{T_c}(t) = \begin{cases}
(v(t) - rw(t)){{k}_{\sigma (t)}}  + {\varpi _c}(t),~~{\rm{if}}\;v(t) \geq w(t)r, ~\sigma (t) \in \mathbb{E} \setminus \mathbb{L}\\
(rw(t) - v(t)){{k}_{\sigma (t)}}  + {\varpi _c}(t),~~{\rm{if}}\;v(t) < w(t)r, ~\sigma (t) \in \mathbb{E} \setminus \mathbb{L}
\end{cases}\nonumber
\end{align}
substituting which together with \eqref{viscoustorque}, \eqref{aerodynamicforce} and \eqref{pressure} into \eqref{carmodel} yields a vehicle model with uncertainties:
\begin{itemize}
  \item if $v(t) \geq w(t)r$, $\sigma (t) \in \mathbb{E} \setminus \mathbb{L}$
  \begin{subequations}
  \begin{align}
  \dot w(t) &= \frac{{{rk_{\sigma (t)}} - \varrho }}{J}w(t) - \frac{{{k_{\sigma (t)}}}}{{J}}v(t) + {u}( t ) + {f_w}( t ),\\
  \dot v(t) &=  - \frac{{{k_{\sigma (t)}}}}{{m}}w(t) + \frac{{{k_{\sigma (t)}} - \zeta {r}}}{{m{r}}}v(t) + {f_v}( t );
  \end{align}\label{newfrom1}
  \end{subequations}
  \item if $v(t) < w(t)r$, $\sigma (t) \in \mathbb{E} \setminus \mathbb{L}$
  \begin{subequations}
  \begin{align}
  \dot{w}(t) &=  - \frac{{{rk_{\sigma (t)}} + \varrho }}{J}w(t) + \frac{{{k_{\sigma (t)}}}}{{J}}v(t) + {u}(t) + {f_w}(t),\\
  \dot{v}(t) &= \frac{{{k_{\sigma (t)}}}}{{m}}w(t) - \frac{{{k_{\sigma (t)}} + \zeta {r}}}{{m{r}}}v(t) + {f_v}(t);
\end{align}\label{newfrom2}
\end{subequations}
\end{itemize}
where ${u}(t)$ denotes control input (${u}(t) > 0$ and ${u}(t) < 0$ indicate activated drive and brake
models, respectively), and
\begin{align}
\!\!\!\!{u}(t) = \frac{{{T_e}(t) - C{\eta _b}{r_b}P(t)}}{J}, ~{f_v}( t ) = \frac{{{\varpi _c}(t) + r\tilde{f}_{v}(t)}}{{mr}},~
{f_w}(t) =  - \frac{{{\varpi _c}(t) + {\varpi _b}(t) + \tilde{f}_{w}(t)}}{J}.\label{aadef}
\end{align}

\section{Safety Envelopes}
This paper considers velocity regulation via tracking the provided references of longitudinal and angular velocities, denoted by  $\mathbf{v}^\mathrm{r}_\sigma$ and $\mathbf{w}^\mathrm{r}_\sigma$, respectively. For the Safe Objective \ref{obj1}, the provided velocity references are required to satisfy the following condition:
\begin{align}
\left|\mathbf{v}^\mathrm{r}_{\sigma(t)} - r\mathbf{w}^\mathrm{r}_{\sigma(t)}\right| - \mathfrak{a}_{\sigma (t)} = 0, \hspace{0.8cm}{\mathfrak{a}_{\sigma (t)}} = \left\{ \begin{gathered}
  \frac{{\zeta r\mathrm{v}_{\sigma (t)}^{\text{r}}}}{{{k_{\sigma (t)}}}},  \hspace{0.2cm}\sigma (t) \in \mathbb{E} \setminus \mathbb{L} \hfill \\
  {\breve{\mathfrak{a}}_{\sigma (t)}},    \hspace{0.7cm}\sigma (t) \in \mathbb{L} \hfill \\
\end{gathered}  \right.\label{referencerelation}
\end{align}
where $\mathbb{E}$ and $\mathbb{L}$ are defined in \eqref{environmentset} and \eqref{learnset}, respectively.

\begin{rem}
\textcolor[rgb]{0.00,0.00,1.00}{We note that vehicle parameters $\zeta$, $r$ and $k_{\sigma (t)}$ in \eqref{referencerelation} indicate that the velocity and slip references depend on the vehicle model. Therefore, the slip reference ${\breve{\mathfrak{a}}_{\text{learned}}}$ in the unforeseen environments depends on the learned models, which is determined according to \eqref{lref11}, such that we can obtain the tracking error dynamics \eqref{impuslerrdyna}.}
\end{rem}

The relation \eqref{referencerelation}, in conjunction with \eqref{slipdef}, indicates that the slip reference is
\begin{align}
{s_{\sigma(t)}^\mathrm{r}} = \left| {\mathbf{v}^\mathrm{r}_{\sigma(t)} - r\mathbf{w}^\mathrm{r}_{\sigma(t)}} \right| = \mathfrak{a}_{\sigma (t)}, \label{sref}
\end{align}
which depends on the driving environment indexed by ${\sigma(t)}$.

The velocity tracking error vector is obtained as
\begin{align}
{e}(t) = [ {{e_w}(t),~{e_v}(t)} ]^\top = [w(t),~v(t)]^\top - [\mathbf{w}^\mathrm{r}_{\sigma(t)},~\mathbf{v}^\mathrm{r}_{\sigma(t)}]^\top, \label{ted}
\end{align}
considering which and \eqref{referencerelation}, we have
\begin{align}
\left| {rw(t) - v(t)} \right| &= \left| {rw(t) - v(t) - r\mathbf{w}^\mathrm{r}_{\sigma(t)} + \mathbf{v}^\mathrm{r}_{\sigma(t)} + \mathfrak{a}_{\sigma (t)}} \right| \nonumber\\
&= \left| {{e_v(t)} -  {re_w(t)}  - \mathfrak{a}_{\sigma (t)}} \right|, \hspace{2.2cm}\mathbf{v}^\mathrm{r}_{\sigma(t)} < r\mathbf{w}^\mathrm{r}_{\sigma(t)}, \label{sa1}\\
\left| {rw(t) - v(t)} \right| &= \left| {v(t) - rw(t) - \mathbf{v}^\mathrm{r}_{\sigma(t)} + r\mathbf{w}^\mathrm{r}_{\sigma(t)} + \mathfrak{a}_{\sigma (t)}} \right| \nonumber\\
&= \left| {{e_v(t)} - {re_w(t)} + \mathfrak{a}_{\sigma (t)}} \right|,   \hspace{2.2cm} \mathbf{v}^\mathrm{r}_{\sigma(t)} \geq r\mathbf{w}^\mathrm{r}_{\sigma(t)}. \label{sa2}
\end{align}

In addition to velocity regulation for collision avoidance, lane keeping, and other constraints, the wheel slip $s(t)$ defined in \eqref{slipdef} should be below a safety boundary $\mu _{\sigma(t)}$ to prevent slipping and sliding, i.e.,
\begin{align}
s(t) = \left|rw(t) - v(t) \right| \le {\mu_{\sigma(t)}},\label{safetyboundary}
\end{align}
which in light of \eqref{sa1} and \eqref{sa2} can be equivalently expressed as
\begin{align}
&- {\mu_{\sigma(t)}} + \mathfrak{a}_{\sigma(t)} \le {e_v(t)} - r{e_w(t)} \le {\mu_{\sigma(t)}} + \mathfrak{a}_{\sigma (t)}, \hspace{0.3cm}\text{if}~\mathbf{v}^\mathrm{r}_{\sigma(t)} < r\mathbf{w}^\mathrm{r}_{\sigma(t)}, \label{safe2}\\
&- {\mu_{\sigma(t)}} - \mathfrak{a}_{\sigma(t)} \le {e_v(t)} - r{e_w(t)} \le {\mu_{\sigma(t)}} - \mathfrak{a}_{\sigma (t)}, \hspace{0.3cm}\text{if}~\mathbf{v}^\mathrm{r}_{\sigma(t)} \geq r\mathbf{w}^\mathrm{r}_{\sigma(t)}. \label{safe1}
\end{align}

Based on \eqref{safe1} and \eqref{safe2}, we define a set of vectors:
\begin{align}
{\widehat{c}_{\sigma \left( t \right)}} = {\left[- \frac{1}{{{\mu _{\sigma (t)}} - {\mathfrak{a}_{\sigma (t)}}}},\hspace{0.2cm}\frac{1}{{\left( {{\mu _{\sigma (t)}} - {\mathfrak{a}_{\sigma (t)}}} \right)r}} \right]^\top}, \hspace{0.2cm}\sigma(t) \in \mathbb{E}\label{safetdef1ad}
\end{align}
by which we obtain the following lemma regarding safety formula.

\begin{lemma}
The safety condition \eqref{safetyboundary} holds if
\begin{align}
-1 &\leq {\widehat c}^{\top}_{\sigma(t)} e(t)  \le 1, \label{safetdef2ad}\\
0 &\leq \mathfrak{a}_{\sigma (t)} < {\mu_{\sigma(t)}}, \label{zkom1}
\end{align}
where $\mathfrak{a}_{\sigma (t)}$, $e(t)$ and ${\widehat c}_{\sigma(t)}$ are given in \eqref{referencerelation}, \eqref{ted} and \eqref{safetdef1ad}, respectively. \label{safetyeqvad}
\end{lemma}

\begin{proof}
\textcolor[rgb]{0.00,0.07,1.00}{Substituting  \eqref{safetdef1ad} into \eqref{safetdef2ad} yields $- 1 \leq \frac{{{e_v}\left( t \right)}}{{\left( {{\mu _{\sigma (t)}} - {\mathfrak{a}_{\sigma (t)}}} \right)r}} - \frac{{{e_w}\left( t \right)}}{{{\mu _{\sigma (t)}} - {\mathfrak{a}_{\sigma (t)}}}} \leq 1$, which, in conjunction with the condition \eqref{zkom1}, leads to
\begin{align}
- {\mu _\sigma } + \mathfrak{a}_{\sigma (t)} \le \frac{{e_v(t)}}{r} - {e_w(t)} \le {\mu _\sigma } - \mathfrak{a}_{\sigma (t)}.\label{safetdef2sdf}
\end{align}
It straightforwardly follows from ${\mu_{\sigma (t)}} - \mathfrak{a}_{\sigma (t)} \le {\mu _\sigma } + \mathfrak{a}_{\sigma (t)}$ that \eqref{safetdef2sdf} implies \eqref{safe1} and \eqref{safe2}. Moreover, the inequalities \eqref{safe1} and \eqref{safe2} equivalently describe the safety condition \eqref{safetyboundary} via the transformations \eqref{sa1} and \eqref{sa2}. We thus conclude that \eqref{safetyboundary} holds if \eqref{safetdef2ad} and \eqref{zkom1} are satisfied.}
\end{proof}

Building on \eqref{safetdef2ad}, safety constraint set for ideal vehicle models is  defined as follows:
\begin{align}
\Omega_{\sigma(t)} = \left\{ {\left.{e(t) \in {\mathbb{R}^2}} \right|c^{\top}_{\sigma(t)}e(t) \le 1} \right\}, ~\sigma(t) \in \mathbb{E},\label{nsafetyset}
\end{align}
where we define
\begin{align}
c_{\sigma(t)}  = \begin{cases}
\widehat c_{\sigma(t)},&{\rm{if}}\;v(t) \geq rw(t)\\
-\widehat c_{\sigma(t)},&{\rm{if}}\;v(t) < rw(t).
\end{cases}\label{coffe}
\end{align}
In addition, we define the following invariant sets and boundary sets, which will be used to determine the safety envelopes:
\begin{align}
{\Phi_{\sigma(t)}} &= \left\{ {\left. {e(t) \in {\mathbb{R}^2}} \right|{e^\top(t)}{\bar{P}_{\sigma(t)}}e(t) \le 1, {{\bar P}_{\sigma(t)}}  > 0} \right\}, ~\sigma(t) \in \mathbb{E}       \label{inset} \\
\partial{\Phi_{\sigma(t)}} &= \left\{ {\left. {e(t) \in {\mathbb{R}^2}} \right|{e^\top(t)}{\bar{P}_{\sigma(t)}}e(t) = 1, {{\bar P}_{\sigma(t)}}  > 0} \right\}, ~\sigma(t) \in \mathbb{E}.   \label{insetboundary}
\end{align}

The following lemma provides a condition under which $\Phi_{\sigma(t)}$ is a subset of safety set $\Omega_{\sigma(t)}$, which will be used for safe velocity regulation.
\begin{lemma}
Consider the safety sets \eqref{nsafetyset}  and \eqref{inset}. $\Phi_{\sigma(t)}$ $\subseteq$ $\Omega_{\sigma(t)}$ holds, if and only if ${\widehat{c}}^{\top}_{\sigma(t)}\bar{P}^{-1}_{\sigma(t)}{\widehat{c}}_{\sigma(t)} \leq 1$, $\sigma(t) \in \mathbb{E}$, where ${\widehat{c}}_{\sigma(t)}$ is given in \eqref{safetdef1ad}.  \label{safetyeqv}
\end{lemma}

\begin{proof}
\textcolor[rgb]{0.00,0.07,1.00}{It is straightforward to verify from \eqref{coffe} that ${c}^{\top}_{\sigma(t)}\bar{P}^{-1}_{\sigma(t)a}{c}_{\sigma(t)} = \widehat{c}^{\top}_{\sigma(t)}\bar{P}^{-1}_{\sigma(t)}\widehat{c}_{\sigma(t)}$. Then, the rest of the proof is  the same as that of Lemma 4.1 in \cite{seto1999engineering} (through letting $x  = e(t)$, $P = \bar{P}^{-1}_{\sigma(t)}$ and $\alpha_{k} = c_{\sigma(t)}$); here it is omitted.}
\end{proof}

In light of Lemma \ref{safetyeqv}, the safety invariant set \eqref{inset} and the safety boundary set \eqref{insetboundary}, we present the safety envelopes for vehicles driving in different environments:
\begin{align}
\text{\underline{Safety Envelopes}:}~~~~~~~~~~~~~~~~~~~ \Theta_{\sigma} = \{ {\left. {e \in {\mathbb{R}^2}} \right|{e^\top}{\bar{P}{_{\sigma)}}}e \le \theta ~\text{and} \mathop {\min }\limits_{y \in \partial {\Phi{_{\sigma}}}} \left\| {e - y} \right\| \ge {\varepsilon}} \},\label{safenvelop}
\end{align}
where $0 < \theta < 1$ and $0 < \varepsilon < 1$.

\section{Model Switching and Model Learning}
As shown in Fig. \ref{RHAC}, the operation of M$\mathcal{L}_{1}$HAC relies on the off-line-built vehicle models corresponding to the prepared environments and the
learned models from finite-time model leaning in an unforeseen driving environment.

\subsection{Off-Line-Built Switching Models}
The stored off-line-built switching models are straightforwardly obtained from \eqref{newfrom1} and \eqref{newfrom2} via dropping the uncertainties as follows:
\begin{itemize}
  \item if $v(t) \geq rw(t)$,
  \begin{align}
\hspace{0.30cm}\dot{\bar{\mathrm{w}}}(t) = \frac{{{rk_{\sigma (t)}} - \varrho }}{J}\bar{\mathrm{w}}(t) - \frac{{{k_{\sigma (t)}}}}{{J}}\bar{\mathrm{v}}(t) + \bar{\mathrm{u}}(t),
\hspace{0.45cm}\dot{\bar{\mathrm{v}}}(t) = - \frac{{{k_{\sigma (t)}}}}{{m}}\bar{\mathrm{w}}(t) + \frac{{{k_{\sigma (t)}} - \zeta {r}}}{{m{r}}}\bar{\mathrm{v}}(t);\label{idnewfrom1}
\end{align}
\item if $v(t) < rw(t)$,
\begin{align}
\dot{\bar{\mathrm{w}}}(t) =  - \frac{{{rk_{\sigma (t)}} + \varrho }}{J}\bar{\mathrm{w}}(t) + \frac{{{k_{\sigma (t)}}}}{{J}}\bar{\mathrm{v}}(t) + \bar{\mathrm{u}}(t), \hspace{0.2cm}\dot{\bar{\mathrm{v}}}(t) = \frac{{{k_{\sigma (t)}}}}{{m}}\bar{\mathrm{w}}(t) - \frac{{{k_{\sigma (t)}} + \zeta {r}}}{{m{r}}}\bar{\mathrm{v}}(t).\label{idnewfrom2}
\end{align}
\end{itemize}

We now define
\begin{align}
\bar{\mathrm{x}}(t) \!=\!\! \left[ \begin{gathered}
  \bar{\mathrm{w}}(t) \hfill \\
  \bar{\mathrm{v}}(t) \hfill \\
\end{gathered}  \right]\!\!,~B \!=\!\! \left[ \begin{gathered}
  1 \hfill \\
  0 \hfill \\
\end{gathered}  \right]\!\!, ~{A_{\sigma_{1}(t)}} \!=\!\! \left[\!\!\! {\begin{array}{*{20}{c}}
  {\frac{{{rk_{\sigma(t)} - \varrho}}}{J}} \!\!&\!\! { - \frac{{{k_{\sigma(t)}}}}{{J}}} \\
  { - \frac{{{k_{\sigma(t)}}}}{{m}}} \!\!&\!\! {\frac{{{k_{\sigma(t)} - r\zeta}}}{{m{r}}}}
\end{array}} \!\!\!\right]\!\!, ~{A_{\sigma_{2}(t)}} \!=\!\! \left[\!\!\! {\begin{array}{*{20}{c}}
  { - \frac{{{rk_{\sigma(t)} + \varrho }}}{J}} \!\!&\!\! {\frac{{{k_{\sigma(t)} }}}{{J}}} \\
  {\frac{{{k_{\sigma(t)} }}}{{m}}} \!\!&\!\! { - \frac{{{k_{\sigma(t)} + r\zeta}}}{{m{r}}}}
\end{array}} \!\!\!\right]\!\!,\label{orgmatrix}
\end{align}
by which, the off-line-built switching models, consisting of \eqref{idnewfrom1} and \eqref{idnewfrom2}, are rewritten as
\begin{align}
\dot{\bar{\mathrm{x}}}( t ) &= {A_{\widetilde{\sigma}(t)}}\bar{\mathrm{x}}( t ) + B\bar{\mathrm{u}}( t ), \hspace{0.50cm}\textcolor[rgb]{0.00,0.00,1.00}{\widetilde{\sigma}(t)  = \begin{cases}
\sigma_{1}(t),&v(t) \geq w(t)r, \sigma (t) \in \mathbb{E} \setminus \mathbb{L}\\
\sigma_{2}(t),&v(t) < w(t)r, \sigma (t) \in \mathbb{E} \setminus \mathbb{L}\\
\sigma(t) \in \mathbb{L}. \\
\end{cases}}\label{reideal}
\end{align}
Meanwhile, the real vehicle dynamics described by \eqref{newfrom1} and \eqref{newfrom2} is rewritten as
\begin{align}
\dot x( t ) &= {A_{\widetilde{\sigma}(t)}}x( t ) + Bu( t ) + f_{0}(x, t ),\label{redynana}
\end{align}
where $f_{0}(x, t ) = \left[ {f_{w}(t),f_{v}(t)} \right]^\top$.

\begin{rem}
We handle the un-modeled forces/torques, e.g.,  rolling resistance forces, as uncertainties, which will be compensated by $\mathcal{L}_{1}$ adaptive controller in M$\mathcal{L}_{1}$HAC.
\end{rem}

\subsection{Finite-Time Model Learning}
\subsubsection{Model Learning Procedure}
The unknown and unmeasured environmental characteristics can potentially lead to large mismatch between the off-line-built vehicle-environment interaction model \eqref{reideal} and real vehicle behaviors in  unforeseen environments. Subsequently, the control action cannot be reliable. This motivates to employ finite-time model learning to timely learn and update a vehicle model using the most recent sensor data generated in the unforeseen environment.

Without loss of generality, the real vehicle dynamics in an unforeseen environment is written as
\begin{align}
\dot x( t ) &= {A_{\text{learned}}}x( t ) + B_{\text{learned}}\widehat{u}( t ) + f_{1}(x,t), ~~~~~~~~~~~~~~~~~\text{learned} \in \mathbb{L} \label{redyna}
\end{align}
where $x(t) = [w(t),~v(t)]^{\top}$, $B_{\text{learned}} = [b_{\text{learned}},~0]^{\top}$, $\widehat{u}( t ) \in \mathbb{R}$ is the control input, and $f_{1}(x,t)$ denotes the uncertainty. The data sampling technique transforms the continuous-time dynamics \eqref{redyna} to the discrete-time one:
\begin{align}
x(q + 1) = ( {\mathbf{I} + T{A_{\text{learned}}}})x( q ) + T B_{\text{learned}}\widehat{u}( q ) + Tf_{1}(x,q),  \hspace{0.6cm}y( {q}) = x( {q}) + \mathbf{o}(q), \label{kkm}
\end{align}
where $T$ is the sampling period, $y\left( {q} \right)$ is the observed sensor data, $\mathbf{o}\left( q\right)$ is the observation/sensing noise, $q \in \{k, k+1, \ldots, k+m \}$, and \begin{align}
k = \frac{{{t - \kappa}}}{T},\hspace{1.0cm}m = \frac{\kappa}{T}. \label{disckk}
\end{align}

\begin{rem}
\textcolor[rgb]{0.00,0.07,1.00}{The $m$ in \eqref{disckk} denotes the number of collected sensor data. It follows from \eqref{disckk} that $kT = t-\kappa$ and $( {k + m})T = t$, which indicates that the state samplings in the time interval $[t-\kappa, t]$ are used to learn the vehicle model denoted by $({A_{\text{learned}}}, {B_{\text{learned}}})$.}
\end{rem}

Moving forward, we introduce:
\begin{align}
\widehat{A} &= {\mathbf{I} + T{A_{\text{learned}}}}, \hspace{1.0cm}\widehat{B} = T B_{\text{learned}}, \hspace{1.0cm}\widehat{u}( p ) \equiv \mathfrak{u}, \forall p \in \{k, k+1, \ldots, k+m \} \label{deino}
\end{align}
where the third term in \eqref{deino} means that the control input keeps constant for the sake of learning.

With the consideration of \eqref{deino}, following the finite-time learning procedure developed in \cite{mlearn} we have
\begin{align}
\widehat{A}_{\text{learned}} = \breve{Q}{\breve{P}^{ - 1}}, \hspace{1.0cm}\widehat{B}_{\text{learned}}\mathfrak{u} = \frac{1}{m}\sum\limits_{z = k}^{k+m-1} {\left( {y\left( {z + 1} \right) - {{\widehat{A}}_{\text{learned}}}y\left( z \right)} \right)}  \label{discle}
\end{align}
where $\widehat{A}_{\text{learned}}$ and $\widehat{B}_{\text{learned}}$ denote the learned ones corresponding to $\widehat{A}$ and $\widehat{B}$, and
\begin{align}
  \breve{P} = \sum\limits_{p = k}^{p - 2} {\sum\limits_{p < q}^{p-1} {\mathbf{y}_p^q{{\left( {\mathbf{y}_p^q} \right)^\top}}}}, \hspace{1.0cm}\breve{Q} = \sum\limits_{p = k}^{p - 2} {\sum\limits_{p < q}^{p-1} {\mathbf{y}_{p + 1}^{q+1}{{\left( {\mathbf{y}_p^q} \right)^\top}}} }, \hspace{1.0cm} \mathbf{y}^{q}_{p} = y( p ) - y( q ). \nonumber
\end{align}
Recalling \eqref{deino} and \eqref{discle}, the learned model is obtained as
\begin{subequations}
\begin{align}
{\breve{A}_{{\text{learned}}}} &= \frac{1}{T}( {{{\widehat A}_{{\text{learned}}}} - \mathbf{I}} ) = \frac{1}{T}( {\breve{Q}{\breve{P}^{ - 1}} - \mathbf{I}}),\\
{\breve{B}_{{\text{learned}}}} &= \frac{\sum\limits_{z = k}^{k+m-1} {( {y(z + 1) - {{\widehat A}_{{\text{learned}}}}y(z)})}}{{mT\mathfrak{u}}}  = \frac{\sum\limits_{z = k}^{k+m-1} {( {y(z + 1) - \breve{Q}{\breve{P}^{ - 1}}y(z)})}}{{mT\mathfrak{u}}}.
\end{align}\label{akkm}
\end{subequations}

The learned vehicle model in an unforeseen environment for $\mathcal{L}_{1}$ adaptive controller is thus described as
\begin{align}
\dot{\mathrm{x}}( t ) &= {\breve{A}_{\text{learned}}}\mathrm{x}( t ) + \breve{B}_{\text{learned}}\bar{\mathrm{u}}( t ), \hspace{0.20cm}\text{with}~\breve{A}_{\text{learned}} = \left[ {\begin{array}{*{20}{c}}
  {a_{\text{learned}}^{11}}&{a_{\text{learned}}^{12}} \\
  {a_{\text{learned}}^{21}}&{a_{\text{learned}}^{22}}
\end{array}} \right].\label{lref}
\end{align}

With the consideration of the relation \eqref{referencerelation} and the learned $A_{\text{learned}}$ in \eqref{lref}, the chosen velocity and slide references for safe velocity regulation are required to satisfy
\begin{align}
{a_{\text{learned}}^{21}}\mathbf{w}^\mathrm{r}_{\text{learned}} + {a_{\text{learned}}^{22}}\mathbf{v}^\mathrm{r}_{\text{learned}} = 0,\hspace{1.50cm}\left| {r\mathbf{w}^\mathrm{r}_{\text{learned}} - \mathbf{v}^\mathrm{r}_{\text{learned}}} \right| - \breve{\mathfrak{a}}_{\text{learned}} = 0.  \label{lref11}
\end{align}

\subsubsection{Sample Complexity} Due to  modeling uncertainty and sampling noise, one intuitive question pertaining to the accuracy of model learning arises: \emph{given the sampling frequency, how many samplings are sufficient for the learned model to achieve the prescribed levels of accuracy and confidence?} To answer the question, we present the sample complexity analysis of the proposed model learning.

We let $\mathfrak{s}_{i}\left(A_{\text{learned}}\right)$ denote the $i$th singular value of matrix $A_{\text{learned}}$, based on which we assume the following bounds pertaining to $\mathfrak{s}_{i}\left(A_{\text{learned}}\right)$ are known:
\begin{align}
\widehat{\underline{\mathfrak{s}}} &\le \mathop {\min }\left\{ {{\mathfrak{s}_1}\left( A_{\text{learned}} \right), {\mathfrak{s}_2}\left( A_{\text{learned}} \right)} \right\},\nonumber\\
\widetilde{\underline{\mathfrak{s}}} &\le \mathop {\min }\limits_{z \in \left\{ {k + 1, \ldots ,k+m} \right\}} \left\{ {{{\left| {\mathfrak{s}_1^{z - k}\left( A_{\text{learned}} \right) - 1} \right|}}}, {{{\left| {\mathfrak{s}_2^{z - k}\left( A_{\text{learned}} \right) - 1} \right|}}} \right\},\nonumber\\
\overline{\widetilde{\mathfrak{s}}} &\geq \left\| A_{\text{learned}} \right\|_{\mathrm{F}}, \hspace{2.5cm} {\overline{\widehat{\mathfrak{s}}}}_{A_{\text{learned}}} \geq \mathop {\max }\limits_{z \in \left\{ {k + 1, \ldots ,p} \right\}} \left\{ {{{\left\| {{A_{\text{learned}}^{k - 1}} - {A_{\text{learned}}^{z - 1}}} \right\|_{\mathrm{F}}}}} \right\}.\nonumber
\end{align}
With the practical knowledge at hand, the sample complexity analysis is formally presented in the following theorem.
\begin{theorem}\cite{mlearn}
For any $\varepsilon \in [0, 1)$, and any $\rho, \delta \in (0, 1)$, and any $\phi > 0$,
we have: $\mathbf{P} (||\breve{A}_{\mathrm{learned}} - {A}_{\mathrm{learned}}|| \leq \phi) \geq 1 - \delta$, as long as the following hold:
\begin{align}
&\min \left\{ {\frac{{{(1 - \varepsilon )^2\rho^2}}}{{\mathfrak{n}\mathfrak{p}^{2}|| {{{\mathcal{C}}_\mathrm{v}}} ||}},\frac{{(1 - 2\varepsilon )\rho}}{{\mathfrak{p}}}} \right\} \ge \frac{{\gamma ^2}}{2}\ln \frac{{4 {( {\frac{2}{\varepsilon } + 1})^2}}}{\delta },\label{fg0}\\
&\phi \geq \sqrt {\frac{{8c{\kappa ^2}}}{{{\left( {1 - \rho } \right)\mathfrak{f}_{2}l_{\mathrm{up}}}}}\ln {\frac{(2+\rho)^{2}}{\delta\rho^{2} }}}, \label{fg1}
\end{align}
where
\begin{align}
&{\mathfrak{n}} = \sum\limits_{z = k}^{k+m-1} {\left( {( {z + 1})(m+k-z) + \frac{(m+k-z)(m+k-z+1)}{2}}\right)},\\
&\eta = [\breve{\mathbf{x}}^{\top}(1), ~~{{\widehat{\bf{w}}}^{\top}}, ~~\breve{\mathbf{f}}^\top,~~  \widetilde{\mathbf{f}}^\top]^{\top}, \hspace{1.0cm} \mathcal{C}_{\mathrm{v}} =  {\bf{E}}[\eta\eta^{\top}], \hspace{1.0cm} \mathfrak{p} = \frac{{{{\sqrt{\mathfrak{n}}\sqrt{2}}{\mathfrak{g}^2}}}}{{\sum\limits_{r = k}^{k+m - 1} {\sum\limits_{q = r+1}^{k+m} {{\mathfrak{f}_{\left( {r,q} \right)}}} } }}, \nonumber\\
&\mathfrak{f}_{(r,q)} = \begin{cases}
\mathfrak{f}_{1}, &\text{if}~q > 2r- 1\\
\mathfrak{f}_{2}, &\text{if}~q  \leq 2r- 1
\end{cases}, \hspace{0.5cm}\mathfrak{f}_{1} = \widehat{\underline{\mathfrak{s}}}^{2k - 2}\widetilde{\underline{\mathfrak{s}}}^{2}\sigma_{\mathrm{i}}^2 + \widehat{\underline{\mathfrak{s}}}^{2k-2}\mathfrak{v}_{\rm{p}}^2 + 2\mathfrak{v}_{\mathrm{o}}^2,  \hspace{0.5cm} \mathfrak{f}_{2} = \widehat{\underline{\mathfrak{s}}}^{2k - 2}\widetilde{\underline{\mathfrak{s}}}^{2} \mathfrak{v}_{\mathrm{i}}^2 \!+\! 2\mathfrak{v}_{\mathrm{o}}^2,\nonumber \\
&\mathfrak{g} = 1 + \overline{\widehat{\mathfrak{s}}} + \mathop {\max }\limits_{q \in \left\{ {k, \ldots ,k+m - 1} \right\}}\left\{ {\frac{{{\overline{\widetilde{\mathfrak{s}}}} - \overline{\widetilde{\mathfrak{s}}}^{q - 1}}}{{1 - {\overline{\widetilde{\mathfrak{s}}}}}}}\right\} + \mathop {\max }\limits_{j < r \in \left\{ {k, \ldots ,k+m - 1} \right\}}\left\{{\frac{{\overline{\widetilde{\mathfrak{s}}}^{j - 1} -\overline{\widetilde{\mathfrak{s}}}^{r - 1}}}{{1 - {\overline{\widetilde{\mathfrak{s}}}}}}}\right\},\nonumber
\end{align}
with $\mathfrak{v}_{\mathrm{p}}^2$, $\mathfrak{v}_{\mathrm{o}}^2$ and $\mathfrak{v}_{\mathrm{i}}^2$ respectively denoting the variances of $Tf_{1}(x,q)$, $\mathbf{o}(q)$ and the initial condition of the dynamics \eqref{kkm}, and
\begin{align}
&\breve{\mathbf{x}}(1) = \left[ {{x^\top}(1),~{x^\top}(1), ~\ldots, ~{x^\top}(1)} \right]\!^{\top} \in {\mathbb{R}^{\sum\limits_{r = k}^{k+m - 1} \!\!{2(k+m-r)} }},\nonumber\\
&{\widehat{\mathbf{w}}}_{k} =  [ {{{(\mathbf{o}(k)-\mathbf{o}(k+1))^\top}}\!, \ldots, {{(\mathbf{o}(k)-\mathbf{o}(k+m))^\top}}} ], \nonumber\\
&{{\widehat{\bf{w}}}} = [ {{\widehat{\bf{w}}}_k,~ {\widehat{\bf{w}}}_{k + 1},~ \ldots,~ {\widehat{\bf{w}}}_{k+m-2},~ {\widehat{\bf{w}}}_{k+m-1}}]^{\top},\nonumber\\
&\breve{\mathbf{f}}^{r}_{k} =  [ {{{(Tf_{1}(x,k-1) - Tf_{r}(x,r-1))^\top}},~ \ldots, ~{{(Tf_{1}(x,r-k+1) - Tf_{r}(x,1))^\top}}}],\nonumber\\
&\breve{\mathbf{f}}_k =  [ {{{{\breve{\mathbf{f}}_{k}^{k+1}}}},~{{{\breve{\mathbf{f}}_{k}^{k+2}}}}, ~\ldots, ~{{{\breve{\mathbf{f}}_k^{k+m-1}}}}}],\nonumber\\
&\breve{\mathbf{f}} = [ {\breve{\mathbf{f}}_k,~\breve{\mathbf{f}}_{k + 1}, ~\ldots ,~\breve{\mathbf{f}}_{p - 2},~\breve{\mathbf{f}}_{p - 1}}]^{\top},\nonumber\\
&\widetilde{\mathbf{f}}_k^r = [ {(Tf_{1}(x,r-k) + T B_{\text{learned}}\mathfrak{u})^\top, \!~\ldots, \!~(Tf_{1}(x,1) + T B_{\text{learned}}\mathfrak{u})^\top}],\nonumber\\
&\widetilde{\mathbf{f}}_k =  [ {\widetilde{\mathbf{f}}_k^{k+1},~\widetilde{\mathbf{f}}_k^{k+2}, ~\ldots, ~\widetilde{\mathbf{f}}_k^{k+m}} ],\nonumber\\
&\widetilde{\mathbf{f}} = [ {\widetilde{\mathbf{f}}_k,~~\widetilde{\mathbf{f}}_{k + 1},~~ \ldots ,~~\widetilde{\mathbf{f}}_{p - 2},~~\widetilde{\mathbf{f}}_{k+m-1}}]^{\top}.\nonumber
\end{align} \label{mmm}
\end{theorem}

\begin{rem}
Due to page limit, we refer readers to \cite{mlearn} for the moore detailed assumptions of Theorem \ref{mmm}. \textcolor[rgb]{0.00,0.00,1.00}{The parameter $\gamma$ in \eqref{fg0} comes from an assumption in \cite{mlearn} that the distribution of entries of the vector $T(f_{1}(x,k) - f_{1}(x,r)) + \mathbf{o}(k+1) - \mathbf{o}(r+1) - A(\mathbf{o}(k)-\mathbf{o}(r))$ is conditionally $\gamma$-sub-Gaussian.}
\end{rem}

\begin{rem}
\textcolor[rgb]{0.00,0.00,1.00}{Our proposed finite-time model learning is mainly used in the safety-critical and time-critical environments for fast online model updating, when the off-line stored models have a large mismatch with the real system due to unforeseen operating environments or black Swan type events. In the challenging environments, before the availability of the learned model, we do not update the (model based) control command, since without the relatively accurate system model, the computed (model based) control command cannot be not regarded as reliable. However, before collecting the most recent data for learning, we must know in advance how many real-time samples from current trajectory generated in the challenging environments are sufficient for the learned model to satisfy the prescribed levels of accuracy and confidence.  With the consideration of \eqref{fg1}, the number of real-time samples, i.e., $m = l_{\mathrm{up}}$, for model learning, should ensure that \eqref{fg0} and \eqref{fg1} hold, such that the prescribed levels of accuracy $\phi$ and confidence $1 - \delta$ of learned model can be guaranteed.}
\end{rem}

\section{S$\mathcal{L}_{1}$-Simplex Architecture Design}

The off-line-built switching models and on-line leaned models constitute a backbone of S$\mathcal{L}_{1}$-Simplex. Sections 3 and 4 have paved the way to the design of S$\mathcal{L}_{1}$-Simplex, which is carried out in this section.  In this section, we first investigate the safe switching control of the off-line-built and on-line leaned models, which will work as the references of vehicle's safe behaviors for the $\mathcal{L}_{1}$ adaptive controller to track.

\subsection{Safe Switching Control}
This section investigates safe switching control of off-line-built and on-line learned vehicle models. The safe velocity regulation control for the off-line-built model \eqref{reideal} and on-line learned model \eqref{lref} is designed as
\begin{align}
\bar{\mathrm{u}}(t) = - F_{\widetilde{\sigma}(t)}^w{\bar{\mathrm{e}}_w}(t) - F_{\widetilde{\sigma}(t)}^v{\bar{\mathrm{e}}_v}(t) + \mathfrak{b}_{\sigma(t)},
\label{input}
\end{align}
where ${\bar{\mathrm{e}}_w}(t) = \mathrm{w}(t) - \mathbf{w}_{\sigma (t)}^{\text{r}}$, ${\bar{\mathrm{e}}_v}(t) = \mathrm{v}(t) - \mathbf{v}_{\sigma (t)}^{\text{r}}$, $\widetilde{\sigma}(t)$ is given in \eqref{reideal}, $F_{{\widetilde{\sigma}}(t)}^w$ and $F_{{\widetilde{\sigma}}(t)}^v$ are the designed control gains, and
\begin{align}
\mathfrak{b}_{\sigma(t)} = \left\{ \begin{gathered}
 \frac{{\zeta \mathbf{v}_{\sigma (t)}^{\text{r}} + \varrho \mathbf{w}_{\sigma (t)}^{\text{r}}}}{J}, \hspace{1.1cm}{{\sigma(t)}}  \in {\mathbb{E}{\setminus} \mathbb{L}} \hfill \\
 a_{\sigma(t)}^{11}\mathbf{w}_{\sigma (t)}^{\text{r}} + a_{\sigma(t)}^{12}\mathbf{v}_{\sigma (t)}^{\text{r}}, \hspace{0.2cm}{\sigma(t)} \in  \mathbb{L}. \hfill \\
\end{gathered}  \right. \label{cckm}
\end{align}
Substituting \eqref{sa1} and \eqref{sa2} under the constraints \eqref{referencerelation} and \eqref{lref11} into the models \eqref{reideal} and \eqref{lref} with the control input \eqref{input} yields the tracking error dynamics:
\begin{align}
\dot{\bar{\mathrm{e}}}(t) = ( {{A_{\widetilde{\sigma}(t)}} + B_{\sigma \left( t \right)}F_{\widetilde{\sigma}(t)}})\bar{\mathrm{e}}(t), \hspace{0.8cm}\bar{\mathrm{e}}( {t_k}) = {E_k}\bar{\mathrm{e}}( {{t^{-}_k}}), \hspace{0.8cm}\bar{\mathrm{e}}( {t_0^ + }) = \bar{\mathrm{e}}( {{t_0}}),\label{impuslerrdyna}
\end{align}
where ${E_k}$ is due to the impulse effect induced by velocity reference switching, and
\begin{align}
{B_{\sigma \left( t \right)}} = \left\{ \begin{array}{l}
\!\! \left[ {\begin{array}{*{20}{c}}
  1&1 \\
  0&0
\end{array}} \right] = \widehat{B},   \hspace{1.2cm}{{\sigma(t)}}  \in {\mathbb{E}{\setminus} \mathbb{L}}\\
\!\!  \left[ {\begin{array}{*{20}{c}}
  a&a \\
  0&0
\end{array}} \right] = \breve{B}_{\sigma(t)},\hspace{0.75cm}{\sigma(t)} \in  \mathbb{L}.
\end{array} \right. \label{rere}
\end{align}

With the defined $\widehat{B}$ and $\breve{B}_{\text{learned}}$ in \eqref{rere}, we present the LMI formula for computing $F_{\breve{\sigma}}$ and $P_{{\sigma}}$ that guarantee safe velocity regulation:
\begin{subequations}
\begin{align}
&{Q_\sigma} > 0,  \hspace{0.3cm}\forall \sigma \in \mathbb{E}\\
&\widehat{c}^\top _\sigma{Q_\sigma}{\widehat{c}_\sigma} \le 1, \hspace{0.3cm}\forall \sigma \in \mathbb{E}\\
&{A_{\sigma_{\upsilon}}}{Q_\sigma} + \widehat{B}{\breve{E}_{\sigma_{\upsilon}}} + {(\! {{A_{\sigma_{\upsilon}}}{Q_\sigma} \!+\! \widehat{B}{\breve{E}_{\sigma_{\upsilon}}}})^\top} < 0, \hspace{0.1cm}\forall \sigma \in {\mathbb{E}{\setminus} \mathbb{L}}, \upsilon \in \{1,2\} \\
&{A_{\sigma}}{Q_{\sigma}} + \breve{B}_{\sigma}{\breve{E}_{\sigma}} + {( {{A_{\sigma}}{Q_{\sigma}} + \breve{B}_{\sigma}{\breve{E}_{\sigma}}})^\top} < 0, \hspace{0.15cm}\forall \sigma \in \mathbb{L}
\end{align}\label{sasa}
\end{subequations}
based on which,  we obtain
\begin{align}
F_{\sigma_{\upsilon}} = {\breve{E}_{\sigma_{\upsilon}}}\bar{P}_{\sigma}, \hspace{0.2cm}\sigma \in {\mathbb{E}{\setminus}\mathbb{L}}, \upsilon \in \{1,2\}; \hspace{0.6cm}F_{\sigma} = {\breve{E}_{\sigma}}\bar{P}_{\sigma}, \sigma \in \mathbb{L}; \hspace{0.6cm}\bar{P}_{\sigma}^{ - 1} = Q_{\sigma},\sigma \in \mathbb{E}. \label{fnsasa}
\end{align}

\begin{figure}[http]
\centering{
\includegraphics[scale=0.33]{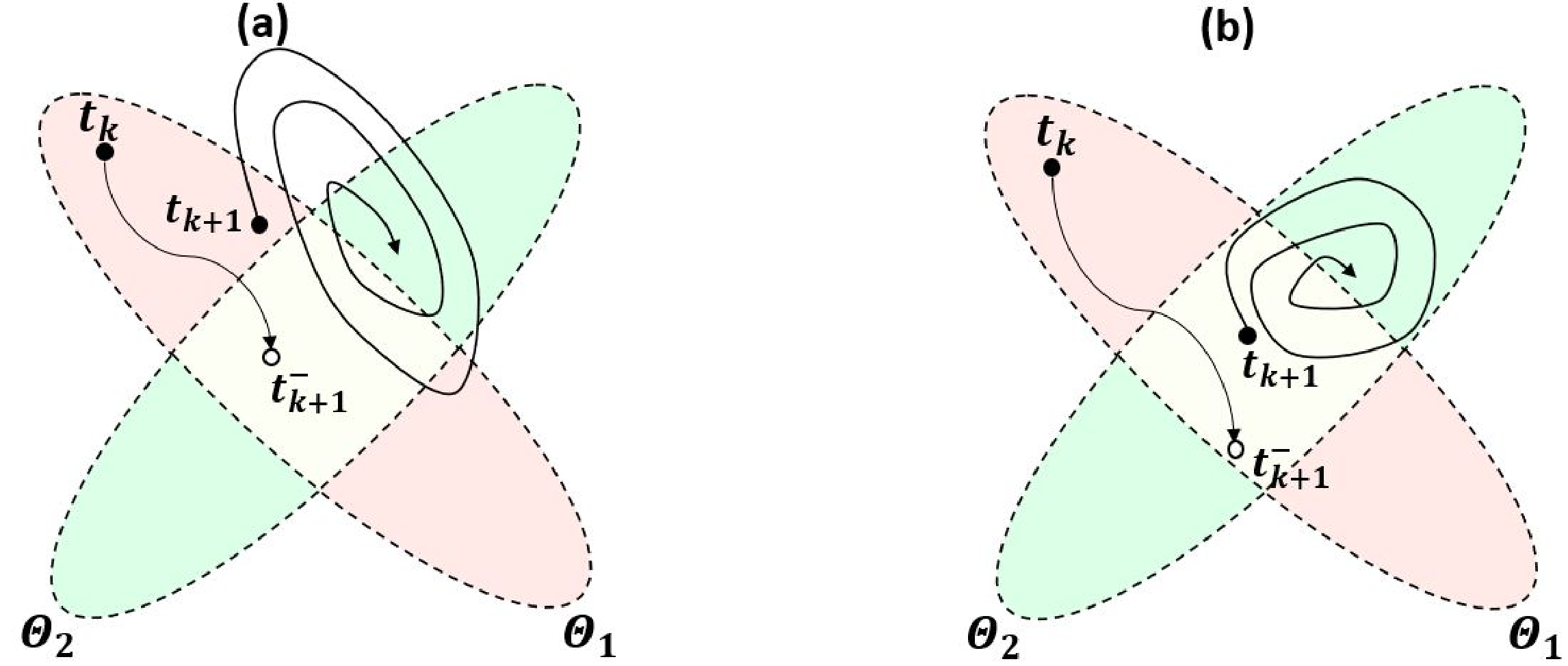}
}
\caption{Impulsive effect on safety envelope: (a) system states escape from safety envelope, (b) system states stay in safety envelope.}
\label{impuls}
\end{figure}

One potential benefit of switching control is extending the safety envelope to $\Theta$  $=$ $\bigcup\limits_{\sigma \in \mathbb{E}}{{\Theta _{\sigma}}}$ \cite{wang2018rsimplex}. However, the switching time and the impulsive effect induced by velocity reference switching are the critical factors in the safety envelope extension. In other words, if the dwell times of subsystems do not take the impulsive effect into account, the safety envelope cannot be extended, which is illustrated by Fig. \ref{impuls}:
\begin{itemize}
  \item Fig. \ref{impuls} (a): at switching time $t_{k+1}$, the system state $\bar{\mathrm{e}}(t_{k+1})$ does not fall into $\Theta_{2}$. Consequently, $\bar{\mathrm{e}}(t) \notin \Theta$ for some time, which hinders the safety envelope extension.
  \item Fig. \ref{impuls} (b): at switching time $t_{k+1}$, the system state $\bar{\mathrm{e}}(t_{k+1})$ falls into the safety envelope $\Theta_{2}$, which leads to $\Theta_{2} \subseteq \Theta$, thus extends the safety envelope.
\end{itemize}
The impulsive effect imposes a higher requirement on the dwell times of switching controllers for the safety envelope extension. For the sake of simplifying presentation of investigation, we define:
\begin{align}
&\bar{A}_{\sigma_{\upsilon}} = {{A_{\sigma_{\upsilon}}} + B{F_{\sigma_{\upsilon}}}}, \hspace{2.50cm}\bar{A}_{\sigma} = {{A_{\sigma}} + \breve{B}_{\sigma}{F_{\sigma}}},\label{smatrix}\\
&\lambda _{\max }^\sigma  = \begin{cases}
\mathop {\max }\limits_{\upsilon \in \left\{ {1,2} \right\}} \{{\lambda _{\max }}({\bar{P}_\sigma }{{\bar A}_{\sigma_{\upsilon}}} + \bar A_{\sigma_{\upsilon}}^\top {\bar{P}_\sigma })\}, &\text{if}~\sigma \in {\mathbb{E}{\setminus}\mathbb{L}}\\
{\lambda _{\max }}({\bar{P}_{\sigma} }{{\bar A}_{\sigma}} + \bar A_{\sigma}^ \top {\bar{P}_{\sigma} }), &\text{if}~\sigma \in \mathbb{L}.
\end{cases}\label{smatrixaddlearn}
\end{align}
With the definitions at hand, the following theorem formally presents safe switching control.
\begin{theorem}
Consider the impulsive switched system \eqref{impuslerrdyna} and safety envelopes \eqref{safenvelop}, with $F_{\sigma_{\upsilon}}$, $F_{\sigma}$  and $\bar{P}_{\sigma}^{ - 1}$ computed via  \eqref{sasa} and \eqref{fnsasa}. If the minimum dwell time defined in \eqref{minimumdwelltime} satisfies
\begin{align}
\mathrm{dwell}_{\min } > \mathop {\max }\limits_{k \in \mathbb{N}} \left\{ {\frac{{{\lambda _{\max }}( {{\bar{P}_{\sigma ({t_{k - 1}})}}})}}{{\lambda _{\max }^{\sigma ({t_{k - 1}})}}}\ln \frac{{{\lambda _{\min }}( {{\bar{P}_{\sigma ({t_{k - 1}})}}})}}{{{\lambda _{\max }}( {E_k^\top{\bar{P}_{\sigma ({t_k})}}{E_k}})}}} \right\},\label{dwellcondiitoncontrolvar}
\end{align}
the system \eqref{impuslerrdyna} is stable, and $\bar{\mathrm{e}}(t) \in \Theta$ for any $t \geq t_{0}$ if $\bar{\mathrm{e}}(t_{0}) \in \Theta_{\sigma(t_{0})}$.\label{safetyimpul}
\end{theorem}

\begin{proof}
In light of \eqref{fnsasa}, the formula \eqref{sasa} equivalently transforms to
\begin{subequations}
\begin{align}
&{\bar{P}_\sigma} > 0, \hspace{0.3cm}\forall \sigma \in \mathbb{E}\label{pth3}\\
&\widehat{c}^\top_\sigma{\bar{P}^{-1}_\sigma}{\widehat{c}_\sigma} \le 1, \hspace{0.3cm}\forall \sigma \in \mathbb{E}\label{pth2}\\
&{\bar{P}_\sigma }{\bar A_{\sigma_{\upsilon}}} + \bar A_{\sigma_{\upsilon}}^\top{\bar{P}_\sigma } < 0, \hspace{0.3cm}\forall \sigma \in {\mathbb{E}{\setminus}\mathbb{L}}, \upsilon \in \{1,2\} \label{pth1}\\
&{\bar{P}_{\sigma}}{\bar A_{\sigma}} + \bar A_{\sigma}^\top{\bar{P}_{\sigma}} < 0, \sigma  \in \mathbb{L} \label{pthk4aa}
\end{align}\label{sasakk}
\end{subequations}
\!\!\!where $\bar{A}_{\sigma_{\upsilon}}$ and ${\bar A_{\sigma}}$ are given in \eqref{smatrix}. We now construct a function:
\begin{align}
{V_{\sigma( {{\bar{t}_k}} )}}( {\bar{\mathrm{e}}( t )} ) = {\bar{\mathrm{e}}^\top}( t ){\bar{P}_{\sigma ( {{\bar{t}_k}} )}}\bar{\mathrm{e}}( t ), \hspace{0.3cm}t \in [ {{\bar{t}_k},{\bar{t}_{k + 1}}}), \label{positive}
\end{align}
where $\bar{t}_{k}$ denotes a switching time. The time derivative of ${V_{\sigma( {{\bar{t}_k}} )}}( {\bar{\mathrm{e}}( t )} ) $ satisfies
\begin{align}
{\dot V_{{\sigma} ({\bar{t}_k})}}(\bar{\mathrm{e}}(t)) &\le \lambda _{\max }^{{\sigma }({\bar{t}_k})}{\bar{\mathrm{e}}^\top}(t)e(t) \label{cdv1}\\
&= \lambda _{\max }^{{\sigma} ({\bar{t}_k})}\lambda _{\max}^{ - 1}( {{\bar{P}_{{\sigma }({\bar{t}_k})}}}){\lambda _{\max }}( {{\bar{P}_{{\sigma} ({\bar{t}_k})}}}){\bar{\mathrm{e}}^\top}(t)\bar{\mathrm{e}}(t) \nonumber\\
& \le \lambda _{\max }^{{\sigma }({\bar{t}_k})}\lambda _{\max }^{ - 1}( {{\bar{P}_{{\sigma }({\bar{t}_k})}}}){V_{{\sigma }({\bar{t}_k})}}(\bar{\mathrm{e}}(t)) < 0,\label{finp1}
\end{align}
where \eqref{cdv1} is obtained via considering \eqref{smatrixaddlearn}, the \eqref{finp1} from its previous step is obtained via considering \eqref{positive}, \eqref{pth1} and \eqref{pthk4aa}.

It follows from \eqref{positive}, \eqref{impuslerrdyna} and \eqref{finp1} that
\begin{align}
{V_{{\sigma} ({\bar{t}_k})}}(\bar{\mathrm{e}}({t_k})) &= {\bar{\mathrm{e}}^{\top}}(\bar{t}_k^ - )({E^{\top}_k{\bar{P}_{{\sigma} ({\bar{t}_k})}}{E_k}})\bar{\mathrm{e}}(\bar{t}_k^ - )\nonumber\\
& \le {\lambda _{\max }}({E_k^{\top}{\bar{P}_{{\sigma} ({\bar{t}_k})}}{E_k}}){\bar{\mathrm{e}}^{\top}}(\bar{t}_k^ - )\bar{\mathrm{e}}(\bar{t}_k^ - )\nonumber\\
& \le \frac{{{\lambda _{\max }}( {E_k^{\top}{\bar{P}_{{\sigma} ({\bar{t}_k})}}{E_k}})}}{{{\lambda _{\min }}( {{\bar{P}_{{\sigma} ({\bar{t}_{k - 1}})}}})}}{V_{{\sigma} ({\bar{t}_{k - 1}})}}(\bar{\mathrm{e}}(\bar{t}_k^ - ))\nonumber\\
& \le \frac{{{\lambda _{\max }}( {E_k^{\top}{\bar{P}_{{\sigma} ({\bar{t}_k})}}{E_k}})}}{{{\lambda _{\min }}( {{\bar{P}_{{\sigma} ({\bar{t}_{k - 1}})}}})}}{e^{\frac{{\lambda _{\max }^{{\sigma} ({\bar{t}_{k - 1}})}( {{\bar{t}_k} - {\bar{t}_{k - 1}}})}}{{{\lambda _{\max }}( {{\bar{P}_{{\sigma} ({\bar{t}_{k - 1}})}}})}}}}{V_{{\sigma }({\bar{t}_{k - 1}})}}(\bar{\mathrm{e}}({\bar{t}_{k - 1}}))\label{finp2ka}\\
& \le \breve{\nu}_{k}{V_{{\sigma }({\bar{t}_{k - 1}})}}\left(\bar{\mathrm{e}}({\bar{t}_{k - 1}})\right),\label{finp2}
\end{align}
where ${\breve{\nu}_k} = \frac{{{\lambda _{\max }}( {E_k^\top{\bar{P}_{\sigma ({\bar{t}_k})}}{E_k}} )}}{{{\lambda _{\min }}( {{\bar{P}_{\sigma ({\bar{t}_{k - 1}})}}} )}}{e^{\frac{{\lambda _{\max }^{\sigma ({\bar{t}_{k - 1}})}\mathrm{dwell}_{\min }}}{{{\lambda _{\max }}( {{\bar{P}_{\sigma ({\bar{t}_{k - 1}})}}})}}}}$. We note that the inequality \eqref{finp2ka} from its previous step is obtained via  considering the integration of \eqref{finp1}, while \eqref{finp2} from \eqref{finp2ka} is obtained via considering $\lambda _{\max }^{^{\sigma ({{\bar t}_{k - 1}})}} < 0$ implied by \eqref{pth1} and \eqref{pthk4aa}. The condition \eqref{dwellcondiitoncontrolvar} implies $0 < \breve{\nu} = \mathop {\min }\limits_{k \in \mathbb{N}} \left\{ {{\breve{\nu}_k}} \right\} < 1$ for $\forall k \in \mathbb{N}$. We thus have
\begin{align}
{V_{{\sigma} ({\bar{t}_k})}}(e({\bar{t}_k})) < \nu{V_{{\sigma} ({\bar{t}_{k - 1}})}}(e({\bar{t}_{k - 1}})), \label{fg}
\end{align}
by which we construct a strictly decreasing sequence with respect to $k$: $\left\{ {{V_{{\sigma} ({\bar{t}_k})}}(\bar{\mathrm{e}}({t_k})),k \in \mathbb{N}} \right\}$. The decreasing sequence straightforwardly implies that the switched system \eqref{impuslerrdyna} is asymptotically stable.

We note that \eqref{finp1} implies that ${V_{{\sigma} ({\bar{t}_k})}}(\bar{\mathrm{e}}({\bar{t}_k})) < {V_{{\sigma} ({\bar{t}_k})}}(\bar{\mathrm{e}}(t))$ for any $t > \bar{t}_k$ in $[ {{\bar{t}_k},{\bar{t}_{k + 1}}})$, which, in conjunction with \eqref{fg}, implies that ${{V}_{\sigma( {{\bar{t}_k}})}}(\bar{\mathrm{e}}(t)) < {{V}_{\sigma( {{\bar{t}_0}})}}(\bar{\mathrm{e}}(t_{0}))$
for any $t > t_{0}$ and $\forall k \in \mathbb{N}$. Therefore, if $\bar{\mathrm{e}}(t_{0}) \in \Theta_{\sigma(t_{0})}$, we have $\bar{\mathrm{e}}(t) \in \Theta_{\sigma(\bar{t}_{0})}$ for any $t \geq t_{0}$. As a consequence, $x(t) \in \Theta = \bigcup\limits_{\sigma \in \mathbb{E}}  {{\Theta_{\sigma}}}$ for any $t \geq t_{0}$. \end{proof}

\begin{rem}
It is known that unreasonable switching between (even) stable models in hybrid systems may lead to instability \cite{liberzon2003switching}. One of the contributions of Theorem \ref{safetyimpul} is to guarantee that under the proposed switching rules, S$\mathcal{L}_{1}$-Simplex will be stable.
\end{rem}

\subsection{Assumption}
We now present a general system than can describe both the vehicle dynamics in the normal environments \eqref{redyna} and the vehicle dynamics in the unforeseen environments \eqref{redynana}:
\begin{align}
\dot x( t ) &= {A_{\widetilde{\sigma}(t)}}x( t ) + B_{\sigma(t)}u( t ) + g(t), \label{redynakkm}
\end{align}
where $\widetilde{\sigma}(t)$ is given in \eqref{reideal}; $B_{{\sigma}(t)}$ is given in \eqref{rere}; $g(t) = f_{0}(x, t)$ if ${\sigma}(t) \in {\mathbb{E}{\setminus}\mathbb{L}}$, and $g(t) = f_{1}(x, t)$, otherwise.

Considering \eqref{redynakkm}, the dynamics of faulty vehicle system can be described by
\begin{align}
\dot x( t ) &= {A_{\widetilde{\sigma}(t)}}x( t ) + B_{\sigma(t)}u( t ) + f_{2}(x, t ),\label{fauredyna}
\end{align}
where  $f_{2}(x, t )$ is an uncertainty function that represents modeling errors, noise, disturbance, unmodeled forces/torques, etc. The fault dynamics \eqref{fauredyna} indicates that this paper focuses on the class of software and physical failures, whose influences can be modeled by $f_{2}(x,t)$.

Building on the safe switching control studied in the previous subsection, the remaining S$\mathcal{L}_{1}$-Simplex design relies on the following assumption on the uncertainty.
\begin{asm}
The uncertainties $f_{q}(x,t)$ in \eqref{redynana}, \eqref{redyna} and \eqref{fauredyna} are uniformly bounded in time and Lipschitz in $x$ over safety set, i.e.,  there exist positive $l_{q}$ and $b_{q}$ such that
\begin{align}
\left\| {f_{q}( {0,t})} \right\| \le b_{q}~\text{and}~\left\| {f_{q}( {{x_1},t}) - f_{q}( {{x_2},t})} \right\| \le l_{q}\left\| {{x_1} - {x_2}} \right\|, ~~q = 0,1,2 \label{assump2eq}
\end{align}
hold for any $t \geq 0$, and $x_{1} - [\mathbf{w}^\mathrm{r}_{\sigma}, \mathbf{v}^\mathrm{r}_{\sigma}]^{\top}$, $x_{2} - [\mathbf{w}^\mathrm{r}_{\sigma}, \mathbf{v}^\mathrm{r}_{\sigma}]^{\top}$ $\in \bigcup\limits_{\sigma  \in \mathbb{E}} {{\Omega_{\sigma}}}$, with $\Omega_{\sigma}$ given in \eqref{nsafetyset}.
\label{assump2gf}
\end{asm}

We next present other backbones of S$\mathcal{L}_{1}$-Simplex in achieving Safe Objective \ref{obj1} and Safe Objective \ref{obj2} simultaneously.

\subsection{Uncertainty Monitor}
As shown in Fig.\ref{RSimplex}, the decision logic needs the measurement of uncertainty from the monitor to make the decision of switching between HPC and M$\mathcal{L}_{1}$HAC. The dynamics of uncertainty monitor of the real car under the control actuator from HPC is described by
\begin{subequations}
\begin{align}
&\dot z( t ) = {A_z}z( t ) + ({A_z}{\widetilde{B}_z} - {\widetilde{B}_z}{A_{\mathrm{hpc}}})x( t ) - {\widetilde{B}_z}Bu( t ),\\
&{\widehat f}(x, t) = {C_z}z( t ) + {C_z}{\widetilde{B}_z}x( t ),\\
&z( {{t_k}} ) =  - {\widetilde{B}_z}x( {{t_k}} ),
\end{align}\label{uncertaintymonitor}
\end{subequations}
\!\!where ${\widehat f}(x, t) \in \mathbb{R}^{2}$ is a measurement of the uncertainty, and the triple $(A_{z}, \widetilde{B}_{z}, C_{z})$ constitutes a low-pass filter \cite{wang2018rsimplex}.

\subsection{Switching Rules}
Building on the safety envelope \eqref{safenvelop} and the uncertainty monitor \eqref{uncertaintymonitor}, the switching rules, including the decision logic for HPC and M$\mathcal{L}_{1}$HAC and the switching logic for off-line-built and learned models, are described below.
\begin{itemize}
  \item Decision Logic: switching \underline{from HPC to M$\mathcal{L}_{1}$HAC}
  \begin{itemize}
  \item \textbf{Rule I:} triggered by the magnitude of uncertainty measurement:
  \begin{align}
  &\left\| {{\widehat f}(x,t)} \right\| > \int_{0}^t {\left\| {{C_z}{e^{{A_z}( {t - \tau })}}{B_z}} \right\|\left( {l_{0}\left\| {x( \tau )} \right\| + {b_{0}}}\right)d\tau}.\label{unmo}
  \end{align}
  \item \textbf{Rule II:} triggered by the safety envelope \eqref{safenvelop}:
  \begin{align}
  {e^\top}( t ){P_\sigma}e( t ) = \theta ~\text{and}~ {e^\top}( t ){P_\sigma}\dot{e}( t ) > 0, \forall \sigma \in \mathbb{E}. \label{save}
  \end{align}
\end{itemize}
  \item Switching Logic: switching \underline{from off-line-built models} to \underline{on-line  learned model}:
  \begin{itemize}
  \item \textbf{Rule III:} triggered by the uncertainty measurement \eqref{unmo} and environmental perception: ${\sigma}( t ) \notin {\mathbb{E}{\setminus}\mathbb{L}}$.
  \item \textbf{Rule IV:} triggered by the safety envelope verification \eqref{save} and environmental perception: ${\sigma}( t ) \notin {\mathbb{E}{\setminus}\mathbb{L}}$.
\end{itemize}
\end{itemize}

\begin{rem}
It has been proved in \cite{wang2018rsimplex} that under Assumption \ref{assump2gf}, i.e., the normal condition, the triggering condition in $\textbf{Rule}$ $\textbf{I}$ does not hold, which means that M$\mathcal{L}_{1}$HAC is not activated.
\end{rem}

\subsection{$\mathcal{L}_{1}$ Adaptive Controller}
The components of $\mathcal{L}_{1}$ adaptive controller  in Fig. \ref{RHAC} are described below.

\subsubsection{State Predictor}
The state predictor of M$\mathcal{L}_{1}$HAC in Figure \ref{RHAC} is described by
\begin{align}
\dot{\tilde x}(t) = {A_{\widetilde{\sigma}(t)}}x( t ) + B_{\sigma(t)}u( t ) + {\tilde{\mathrm{f}}}(t) - \alpha({\tilde x(t) - x(t)}),  \hspace{1.3cm}\tilde x(t_{k^{*}}) = x(t_{k^{*}}),\label{monitor}
\end{align}
where $t_{k^{*}}$ is the switching moment from HPC to M$\mathcal{L}_{1}$HAC, $\alpha$ is an arbitrary positive scalar, and ${\tilde{\mathrm{f}}}(t)$ is the estimation of the uncertainties ${f_0}( {x,t})$, ${f_1}( {x,t})$ and ${f_2}( {x,t})$, which is computed by the following adaptation law.

\subsubsection{Adaptation Law}
The estimated ${\tilde{\mathrm{f}}}(t)$ in \eqref{monitor} is computed via
\begin{align}
{\dot{\tilde{\mathrm{f}}}}(t) = K{\text{Proj}_{{\Psi}}}( {{{\tilde{\mathrm{f}}}}(t), -({\tilde x(t) - x(t)})}),\label{adaptivelaw}
\end{align}
where $K$ is the adaptive gain, and
\begin{align}
\!{\Psi} = \left\{ {f \in {\mathbb{R}^2}\left| {\left\| f \right\| \le \rho  = \frac{l}{{\sqrt {\mathop {\min }\limits_{\sigma  \in \mathbb{E}} \left\{ {{\lambda _{\min }}({P_\sigma })} \right\}} }} + b} \right.} \right\}, \label{PRESET}
\end{align}
with
\begin{align}
l = \max \left\{ {{l_0},{l_1},{l_2}} \right\},\hspace{1.0cm}b = \max \left\{ {{b_0},{b_1},{b_2}} \right\}. \label{bound}
\end{align}

The projection operator ${\text{Proj}_{{\Psi}}}: \mathbb{R}^{2} \times \mathbb{R}^{2} \rightarrow \mathbb{R}^{2}$ in \eqref{adaptivelaw} is defined as
\begin{align}
{\text{Pro}}{{\rm{j}}_{{\Psi}}}( {p,q}) = \left\{ \begin{array}{l}
\!\!q - \frac{{\nabla g( p )(\nabla {g}(p))^\top q g( p )}}{{\left\| {\nabla g( p )} \right\|^{2}}},\hspace{0.3cm}\text{if}~g(p) > 0 ~\text{and}~ q^{\top}\nabla g(p) > 0\\
\!\!q, \hspace{3.4cm}\text{otherwise}
\end{array} \right., \label{PROJECTION}
\end{align}
where $g( p ) = \frac{{{p^\top}p - {\rho ^2} + 1 - \vartheta }}{{1 - \vartheta}}$ with $\vartheta \in ( {1 - {\rho ^2},1} )$. It has been proved in \cite{wang2018rsimplex} that the operator \eqref{PROJECTION} can always guarantee  $\tilde{\mathrm{f}}(t) \in {\Psi}$.

\subsubsection{Low-Pass Filter}
The low-pass filter that takes ${\tilde{\mathrm{f}}}(t)$ as control input is described by
\begin{align}
\dot{\breve{x}}( t ) = \breve{A}\breve{x}( t ) + \breve{B}{{\tilde{\mathrm{f}}}}(t), \hspace{1.8cm}{u^{\text{ad}}}( t ) = \breve{C}\breve{x}( t ), ~~~~~\breve{x}( t^{*}_{k} ) = \mathbf{0},\label{lowpassfilter}
\end{align}
where $t^{*}_{k}$ denotes the switching time when  M$\mathcal{L}_{1}$HAC is activated, the triple $(\breve{A}, \breve{B}, \breve{C})$ is the state space realization of an $1 \times 2$
matrix of low-pass filters that are stable and strictly proper with the transfer function:
\begin{align}
{T}( s ) = \breve{C}{( {s\mathbf{I} - \breve{A}} )^{ - 1}}\breve{B}. \label{transfer}
\end{align}

Finally, as shown in Figure \ref{RHAC}, the control input from M$\mathcal{L}_{1}$HAC for the real car in dynamic and/or unforeseen environments \eqref{redyna} is
\begin{align}
u( t ) = \bar{\mathrm{u}}(t)- {u^{\text{ad}}}( t ), \label{finalcontrol}
\end{align}
where ${u^{\text{ad}}}( t )$ and $\bar{\mathrm{u}}(t)$ are computed by \eqref{lowpassfilter} and \eqref{input}, respectively.

\subsection{M$\mathcal{L}_{1}$HAC Performance}
Finally, we present the performance analysis of M$\mathcal{L}_{1}$HAC. Before proceeding on, we define
\begin{align}
&{H_{\widetilde{\sigma}(t_{k})}}( s ) = {\left( {s\mathbf{I} - {A_{\widetilde{\sigma}(t_{k})}} - B_{\sigma}{F_{\widetilde{\sigma}(t_{k})}}}\right)^{ - 1}}, \label{zzzk1} \\
&\delta  =  {\frac{{{\bar{\mathrm{e}}^ \top }({t_{{k^*}}}){P_{\sigma ({t_{{k^*}}})}}\bar{\mathrm{e}}({t_{{k^*}}})}}{{{\lambda _{\min }}({P_{\sigma ({t_k})}})}}},\label{fth2}\\
&{\varepsilon} = \frac{1}{{(1 - {\chi _{\widetilde{\sigma} ({t_k})}})}}\left( {{{\left\| {{H_{\widetilde{\sigma} ({t_k})}}(s)BT(s)(s + \alpha )} \right\|}_{{\mathcal{L}_1}}}\sqrt {\frac{\mu }{K}} } \right. \left. { + ( {\delta  + \left\| {x_{\sigma({t_k})}^*} \right\| + \frac{b}{l}}){\chi_{\widetilde{\sigma}({t_k})}}} \right),\label{fth2app}\\
&{\chi_{\widetilde{\sigma}(t_{k})}} = {\left\| {{H_{\widetilde{\sigma}(t_{k})}}( s )( {I - BT(s)} )} \right\|_{{\mathcal{L}_1}}}{l}, \label{fth3} \\
&{x_{\sigma(t)}^*} =  [\mathbf{w}^\mathrm{r}_{\sigma(t)},~\mathbf{v}^\mathrm{r}_{\sigma(t)}]^\top. \label{fth345}
\end{align}
With these definitions at hand, the performance of M$\mathcal{L}_{1}$HAC is formally presented in the following theorem.
\begin{theorem}
Consider the real vehicle dynamics \eqref{redynana} with control input \eqref{finalcontrol} from M$\mathcal{L}_{1}$HAC after $t = t_{k^{*}}$. If the minimum dwell time satisfies \eqref{dwellcondiitoncontrolvar}, $e(t_{k^{*}}) \in \Theta_{\sigma(t_{k^{*}})}$,  $\varepsilon > 0$ and ${\varepsilon} + \delta  \le \frac{1}{{\sqrt {{\lambda _{\max }}( {{P_{\sigma ( {{t_k}} )}}} )} }}$, then $e(t) \in \Phi_{\sigma ( {{t_k}} )}$ and $\left\| {x( t ) - \bar{\mathrm{x}}( t )} \right\| \le \varepsilon$, for any $t \in [t_{k},t_{k+1})$, $k \geq k^* \in \mathbb{N}$. \label{final}
\end{theorem}

\begin{proof}
The proof is  similar to the proof path of Theorem 4.10 of \cite{wang2018rsimplex}. We thus only present the critical differences.

It follows from \eqref{redynakkm}, \eqref{reideal} and \eqref{lref}, with the consideration \eqref{input} and \eqref{finalcontrol}, that 
\begin{align}
{\dot{x}( s ) - \dot{\bar{\mathrm{x}}}( s )} = \bar{A}_{\widetilde{\sigma}} ({{x}( s ) - {\bar{\mathrm{x}}}( s )}) - B_{{\sigma}}{u^{\text{ad}}}( t ) + g(t), \nonumber
\end{align}
where $\bar{A}_{\widetilde{\sigma}}$ is given in \eqref{smatrix}. We then have
\begin{align}
{\left\| {x( s ) - \bar{\mathrm{x}}( s )} \right\|_{{\mathcal{L}_\infty }\left[ {{t_k},{t_{k + 1}}} \right)}} \le {\left\| {{H_{\widetilde{\sigma}(t_k)}}(s)BT(s)(s + \alpha )} \right\|_{{\mathcal{L}_1}}}\sqrt {\frac{\mu }{K}} + {{\rm M}_{\widetilde{\sigma}( {{t_k}} )}},\label{np5}
\end{align}
where $H_{\widetilde{\sigma}(t_k)}$ is given in \eqref{zzzk1}, $T(s)$ is given in \eqref{transfer}, and
\begin{align}
\mu  &= 4{\rho ^2} + \frac{{4\alpha {\rho ^2} + 2\rho l}}{\alpha }\left( {\frac{1}{{1 - {e^{ - 2\alpha {\rm{dwell}}{{\rm{l}}_{\min }}}}}} + 1}\right), \label{np5oo1}\\
{{\rm{M}}_{\widetilde{\sigma}(t_k)}} &= {\left\| {{H_{\widetilde{\sigma}(t_k)}}( s )( {\mathbf{I} - BT(s)})} \right\|_{{\mathcal{L}_1}}}{\left\| \breve{f}( s ) \right\|_{{\mathcal{L}_\infty }\left[ {{t_k},{t_{k + 1}}} \right)}}.\label{np5oo2}
\end{align}

Following \eqref{finp1} and \eqref{fg}, we have
\begin{align}
{\bar{\mathrm{e}}^\top}(t){P_{\sigma ({{\bar{t}_k}})}}\bar{\mathrm{e}}(t) = {V_{\sigma( {{\bar{t}_k}})}}( {\bar{\mathrm{e}}( t)}) < {V_{\sigma ( {{t_{{k^*}}}} )}}( {\bar{\mathrm{e}}( {{t_{{k^*}}}})}) = \bar{\mathrm{e}}^\top( {{t_{{k^*}}}}){P_{\sigma ( {{t_{{k^*}}}})}}\bar{\mathrm{e}}( {{t_{{k^*}}}}), \label{hko}
\end{align}
for any $t \in [\bar{t}_{k}, \bar{t}_{k+1})$ with $\bar{t}_{k} \geq t_{k^{*}}$, $\forall k \in \mathbb{N}$. With the consideration of $\delta$ given by \eqref{fth2}, the inequality \eqref{hko} implies that
\begin{align}
{\left\| \bar{\mathrm{e}} \right\|_{{\mathcal{L}_\infty }[{t_k},{t_{k + 1}})}} < \delta, ~~~k \geq k^{*} \in \mathbb{N}.\label{npa1}
\end{align}

It follows from \eqref{assump2eq} that
\begin{align}
{\left\| \breve{f}_{q} \right\|_{{\mathcal{L}_\infty }\left[ {{t_k},{t_{k + 1}}} \right)}} \le {l}{\left\| {x} \right\|_{{\mathcal{L}_\infty }\left[ {{t_k},{t_{k + 1}}} \right)}} + {b}, ~q = 0, 1, 2 \label{unes}
\end{align}
where $l$ and $b$ are given in \eqref{bound}. Combining \eqref{np5oo2} with \eqref{npa1} and \eqref{unes} yields
\begin{align}
{{\rm{M}}_{\widetilde{\sigma}(t_k)}} &\leq {\left\| {{H_{\widetilde{\sigma}({t_k})}}(s)(\mathbf{I} - BT(s))} \right\|_{{\mathcal{L}_1}}}\left( {l{{\left\| {x} \right\|}_{{\mathcal{L}_\infty }[{t_k},{t_{k + 1}})}} + b} \right) \nonumber \\
& = {\chi _{\widetilde{\sigma}({t_k})}}\left( {{{\left\| {x} \right\|}_{{\mathcal{L}_\infty }[{t_k},{t_{k + 1}})}} + \frac{b}{l}} \right)  \nonumber \\
&\leq {\chi _{\widetilde{\sigma}({t_k})}}\left( {{{\left\| {x - \bar{\mathrm{x}}} \right\|}_{{\mathcal{L}_\infty }[{t_k},{t_{k + 1}})}} + {{\left\| { \bar{\mathrm{x}}\left( t \right)} \right\|}_{{\mathcal{L}_\infty }[{t_k},{t_{k + 1}})}} + \frac{b}{l}} \right) \nonumber \\
& \le {\chi _{\widetilde{\sigma}({t_k})}}( {{{\left\| {x - \bar{\mathrm{x}}} \right\|}_{{\mathcal{L}_\infty }[{t_k},{t_{k + 1}})}} + {{\left\| \bar{\mathrm{e}} \right\|}_{{\mathcal{L}_\infty }[{t_k},{t_{k + 1}})}} + \left\| {x_{\widetilde{\sigma}({t_k})}^*} \right\| + \frac{b}{l}})  \nonumber \\
& < {\chi_{\widetilde{\sigma}({t_k})}}( {{{\left\| {x - \bar{\mathrm{x}}} \right\|}_{{\mathcal{L}_\infty }[{t_k},{t_{k + 1}})}} + \delta + \left\| {x_{\sigma ({t_k})}^*} \right\| + \frac{b}{l}}),\label{np4}
\end{align}
where ${\chi _{\widetilde \sigma ({t_k})}}$ is given by \eqref{fth345}.  Substituting \eqref{np4} into \eqref{np5} yields
\begin{align}
\left( {1 \!-\! {\chi _{\widetilde{\sigma}({t_k})}}} \right)\!{\left\| {x \!-\! \bar{\mathrm{x}}} \right\|_{{\mathcal{L}_\infty }[{t_k},{t_{k + 1}})}} \!<\! {\left\| {{H_{\widetilde{\sigma}({t_k})}}\left( s \right)BT\left( s \right)\left( {s \!+\! \alpha } \right)} \right\|_{{\mathcal{L}_1}}}\sqrt {\frac{\mu }{K}}  \!+\! {\chi _{\widetilde{\sigma} ({t_k})}}\!\left( {\delta  \!+\! \left\| {x_{\widetilde{\sigma}({t_k})}^*} \right\| \!+\! \frac{b}{l}} \right),\nonumber
\end{align}
which, in conjunction with $\varepsilon > 0$ (given in \eqref{fth2app}), results in $\left\| {x\left( t \right) - \bar{\mathrm{x}}\left( t \right)} \right\| \le \varepsilon$.
\end{proof}

\section{Experiments}
This section focuses on the demonstration of M$\mathcal{L}_{1}$HAC for safe velocity regulation. The experiments are performed in the AutoRally platform \cite{goldfain2019autorally}, which is a high-performance testbed for self-driving vehicle research. The open source codes of the revised AutoRally platform for the safe velocity regulation in the dynamic and unforeseen environmwents are available at \textcolor[rgb]{1.00,0.00,1.00}{\url{https://github.com/ymao578/GM}}.

\subsection{AutoRally Knowledge}
\subsubsection{Actuators} The throttle, steering and brakes are the control variables of AutoRally.  The valid actuator command values in the steering are between $[-1,1]$. The
steering values of $-1$, 1 and 0 will turn the steering all the way left, all the way right and make any calibrated AutoRally platform drive in a straight line, respectively.

The valid actuator command values in the throttle and front brake are between $[-1,1]$. A throttle value of $-1$ is full (rear) brake and 1 is full throttle. The front brake value ranges from 0 for no brake to 1 for full front brake while negative values are undefined.

\subsubsection{Vehicle Model Parameters} All simulation vehicle parameters (including total mass, front wheel mass, rear wheel mass, overall length, overall width, overall height, wheelbase, rear axle to CG (x offset), rear axle to CG (z offset), front track, rear track and wheel diameter, and sensor placement and characteristics) are set according to their experimentally determined values from a physical 1:5 scale (HPI Baja 5SC) RC trophy truck \cite{goldfain2019autorally}.

The vehicle's parameters, including wheel rotational inertia, friction torque on wheel, aerodynamic drag constant, viscous friction in driven wheel, gravity center height, brake piston effective area, pad friction coefficient and brake disc effective radii, are unknown \cite{goldfain2019autorally}.

\subsubsection{Vehicle Setting}
In the experiments, the vehicle's actuator command value of steering is fixed as 0, i.e., the vehicle is driving straightforwardly. The front brake is disused. The sensor sampling frequency of the angular and longitudinal velocities are set to 100Hz. The driving areas are flat.

\subsection{$\mathcal{L}_{1}$ Adaptive Controller v.s. Normal Controller } 
\begin{figure}[!htb]
\minipage{0.50\textwidth}
  \includegraphics[width=\linewidth]{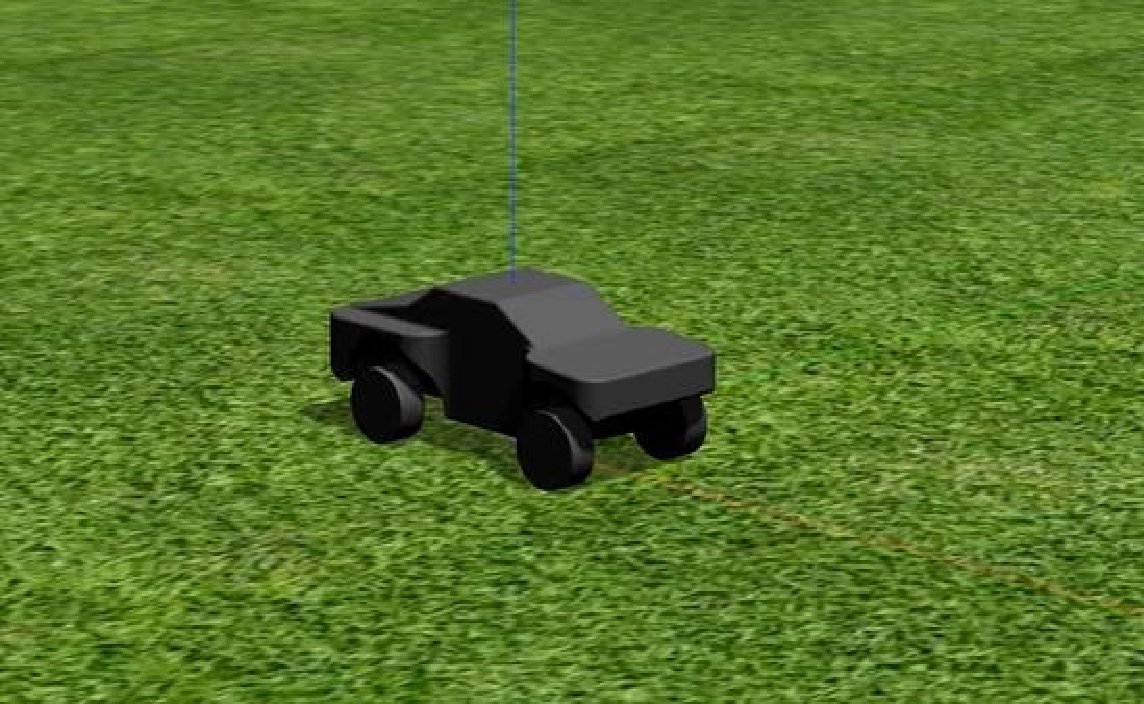}
  \caption{Driving Environment}\label{fig:sa1}
\endminipage\hfill
\minipage{0.50\textwidth}
  \includegraphics[width=\linewidth]{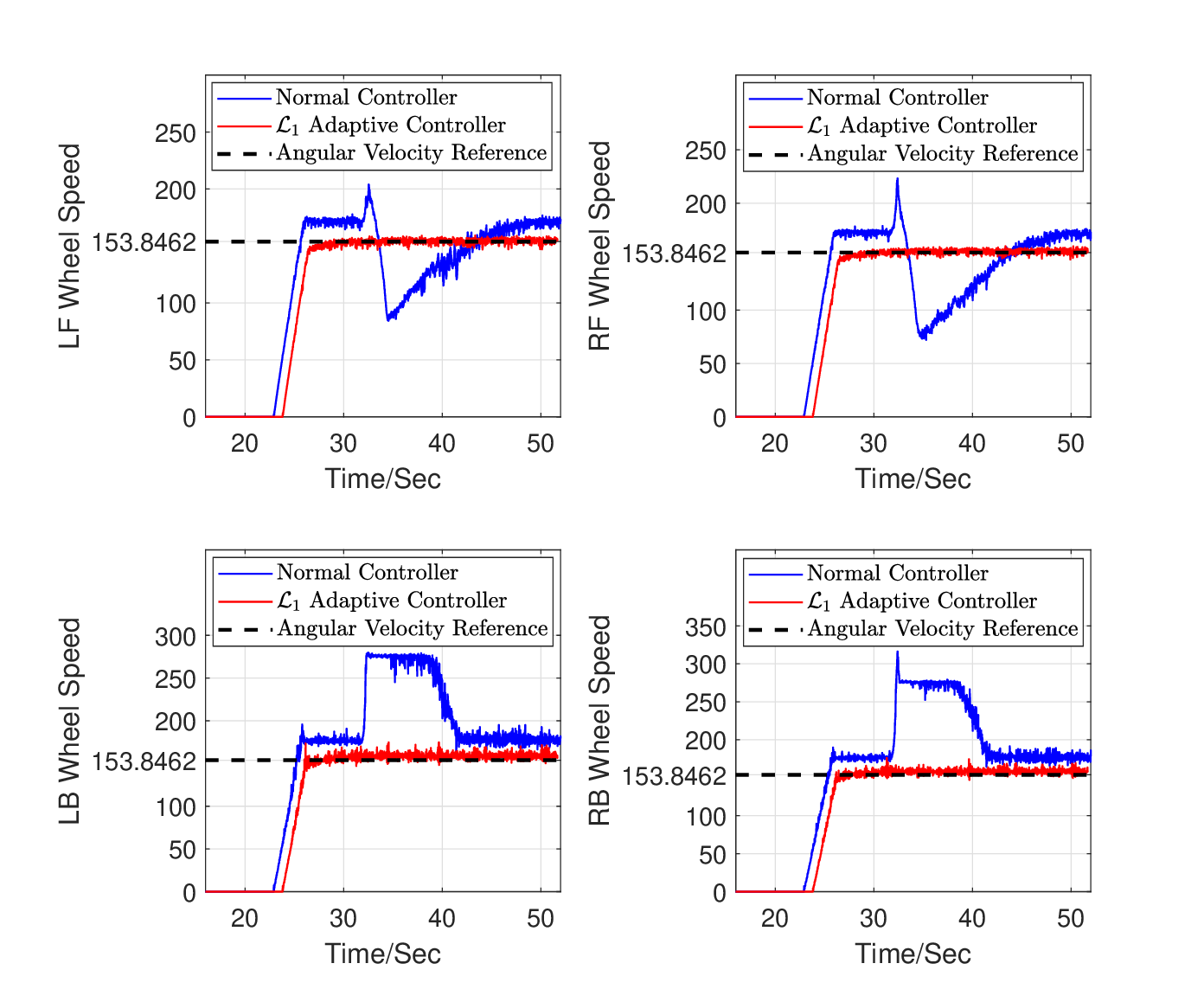}
  \caption{Wheel Angular Velocities}\label{fig:sa2}
\endminipage\hfill
\minipage{0.50\textwidth}
\includegraphics[width=\linewidth]{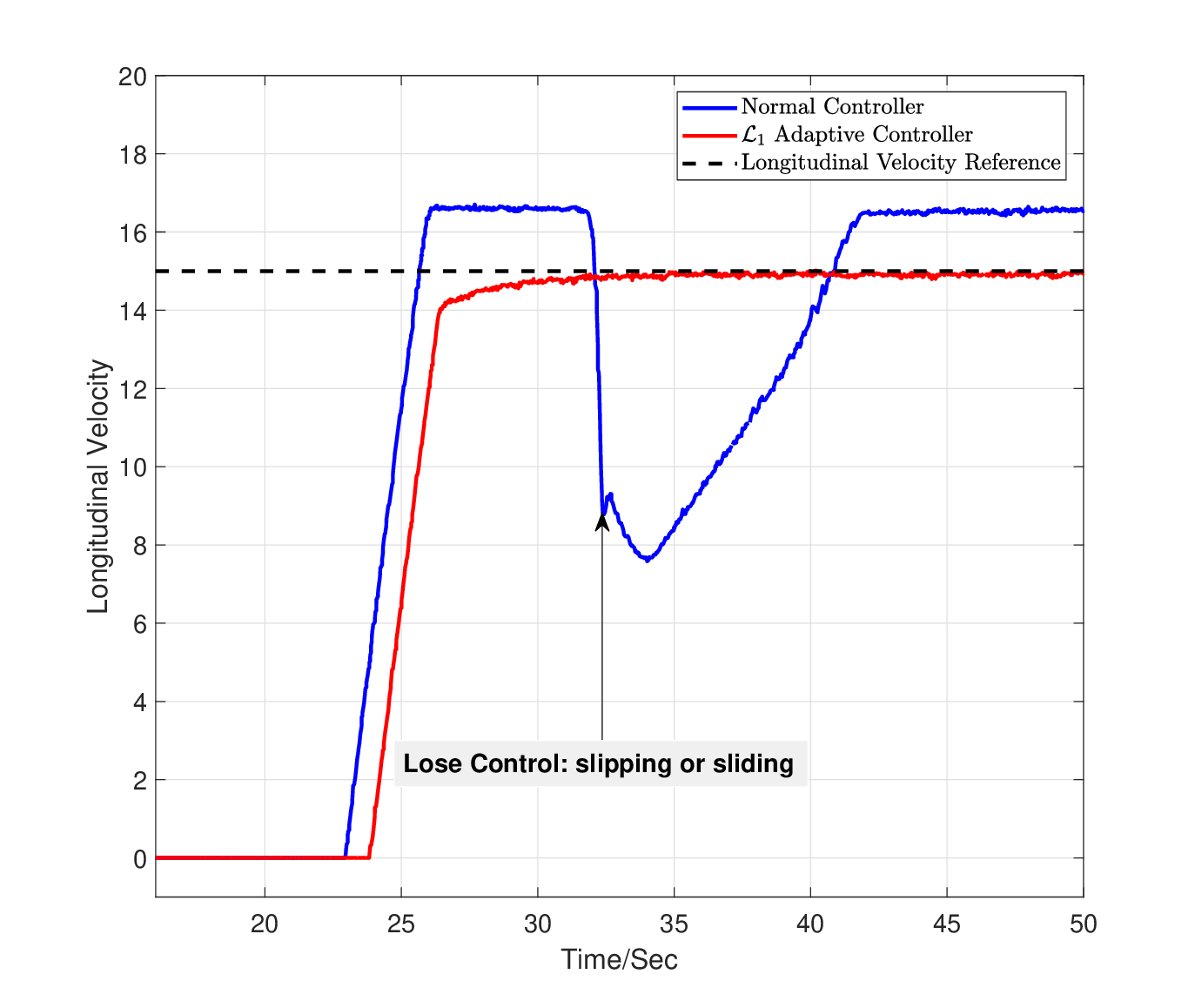}
\caption{Longitudinal Velocities}\label{fig:sa3}
\endminipage\hfill
\minipage{0.50\textwidth}%
  \includegraphics[width=\linewidth]{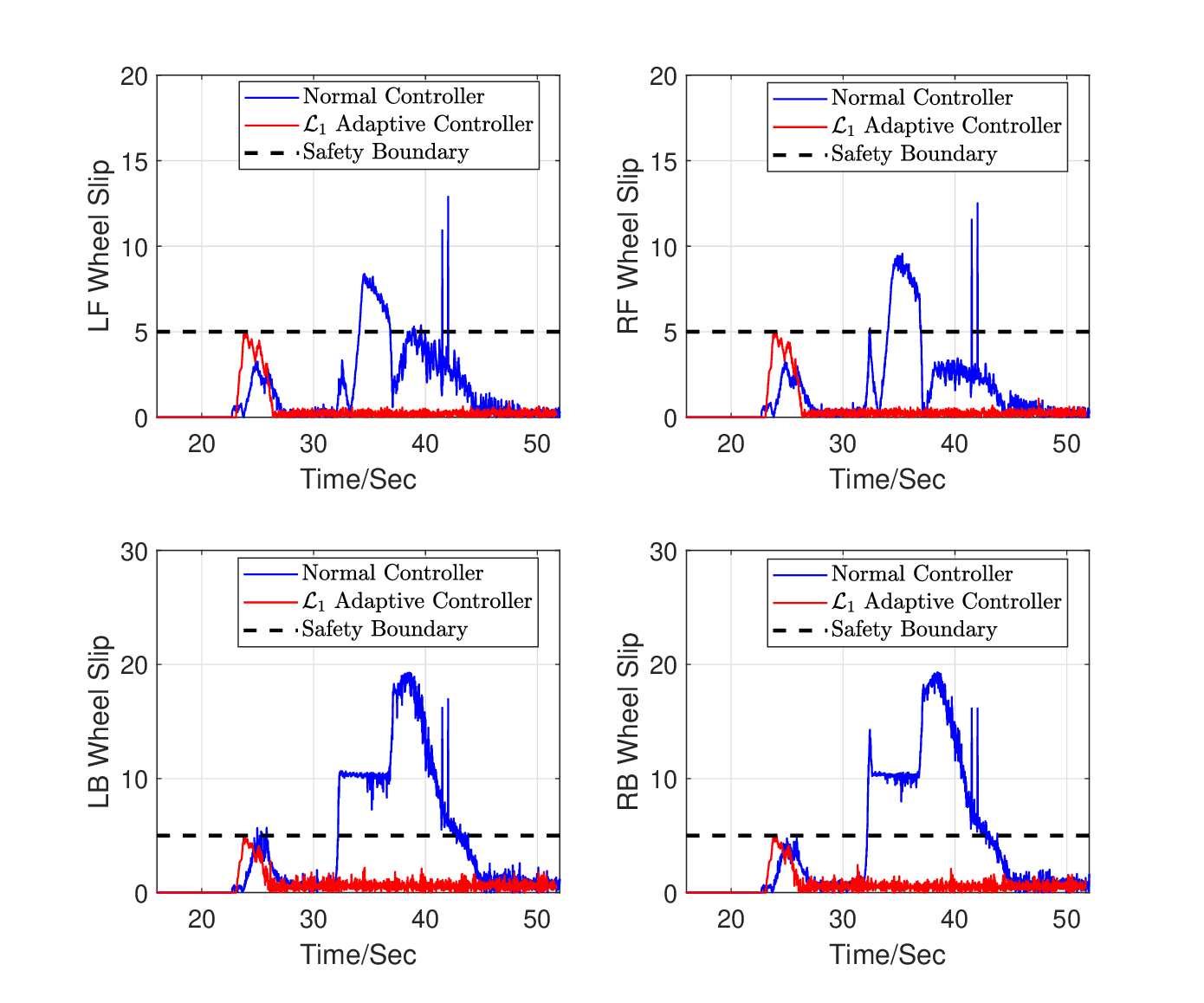}
  \caption{Wheel Slips}\label{fig:sa4}
\endminipage
\end{figure}

Due to the unknown parameters of AutoRally listed in the subsection 6.1.2, the off-line built models described by \eqref{idnewfrom1} and \eqref{idnewfrom2} are not available.  Alternatively, we use the learned model to demonstrate the advantage of $\mathcal{L}_{1}$ adaptive controller. As shown in Fig. \ref{fig:sa1}, the vehicle is driving in a flat grass area. With the computed variances, according to Theorem \ref{mmm}, the sensor data in 3 seconds time interval can guarantee the prescribed levels of accuracy $\phi = 0.8$ and confidence $1 - \delta = 0.8$ of the learned model:
${A_{\text{learned}}} = \left[ {\begin{array}{*{20}{c}}
  {{\text{0}}{\text{.2753}}}&{ - {\text{0}}{\text{.4740}}} \\
  {{\text{1}}{\text{.1742}}}&{ - {\text{1}}{\text{.3313}}}
\end{array}} \right]$ \& ${B_{\text{learned}}} = \left[ {\begin{array}{*{20}{c}}
  {{\text{0}}{\text{.7}}}&{0.7} \\
  {\text{0}}&0
\end{array}} \right]$.

We set the slip safety boundaries as $\mu_{\mathrm{grass}} = 5$m/sec. For the state bias, we let $\varepsilon = 2$. With the knowledge of wheel radius $r =  0.0975$m, we set the references of angular and longitudinal velocities, respectively, as $\mathbf{w}^{r} = 153.8462$rad/sec and  $\mathbf{v}^{r} = 15$m/sec, and the safety boundaries of slip is set to $\mu_{\mathrm{grass}}$ = 5m/sec. We let the minimum dwell time be $\mathrm{dwell}_{\min} =$ 0.5sec. The controller matrix is solved by LMI toolbox as $F_{\text{learned}} = \left[ {\begin{array}{*{20}{c}}
  {{\text{109}}{\text{.8254}}}&{ - {\text{32}}{\text{.5282}}} \\
  {{\text{109}}{\text{.3218}}}&{ - {\text{33}}{\text{.5638}}}
\end{array}} \right]$. For $\mathcal{L}_{1}$ adaptive controller, we set the adaptive law parameters as $K = 10$, $\rho = 1$ and $\upsilon = 0.5$. We set the low-pass filter matrices as $\breve{A} = \breve{B} = \breve{C} = \left[ {\begin{array}{*{20}{c}}
  1&1 \\
  0&1
\end{array}} \right]$. The state predictor gain parameter is set to $\alpha = 5$.

The trajectories of angular velocities, longitudinal velocities and slip are respectively shown in Figs. \ref{fig:sa2}-\ref{fig:sa4}, from which we observe that
\begin{itemize}
  \item $\mathcal{L}_{1}$ adaptive controller succeeds in achieving safe velocity regulation, i.e., the vehicle's angular and longitudinal velocities successfully track their references and the four wheel slips are always below the safety boundary;
  \item using the normal controller (i.e., only the normal control input \eqref{input}), the vehicle cannot achieve the safe velocity regulation and finally loses control.
\end{itemize}
The demonstration video is available at: \textcolor[rgb]{1.00,0.00,1.00}{\url{https://ymao578.github.io/pubs/m2.mp4}}.

\subsection{M$\mathcal{L}_{1}$HAC  v.s. $\mathcal{L}_{1}$HAC}
\begin{figure}[!htb]
\minipage{0.50\textwidth}
  \includegraphics[width=\linewidth]{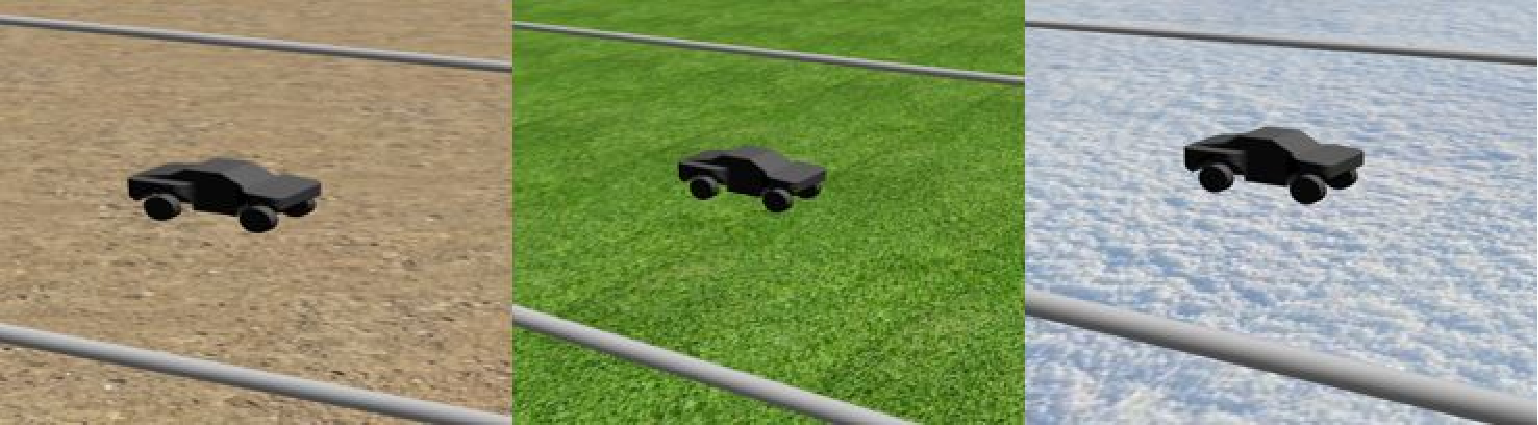}
  \caption{Dynamic and Unforeseen Driving Environemnts}\label{fig:ssa1}
\endminipage\hfill
\minipage{0.50\textwidth}
  \includegraphics[width=\linewidth]{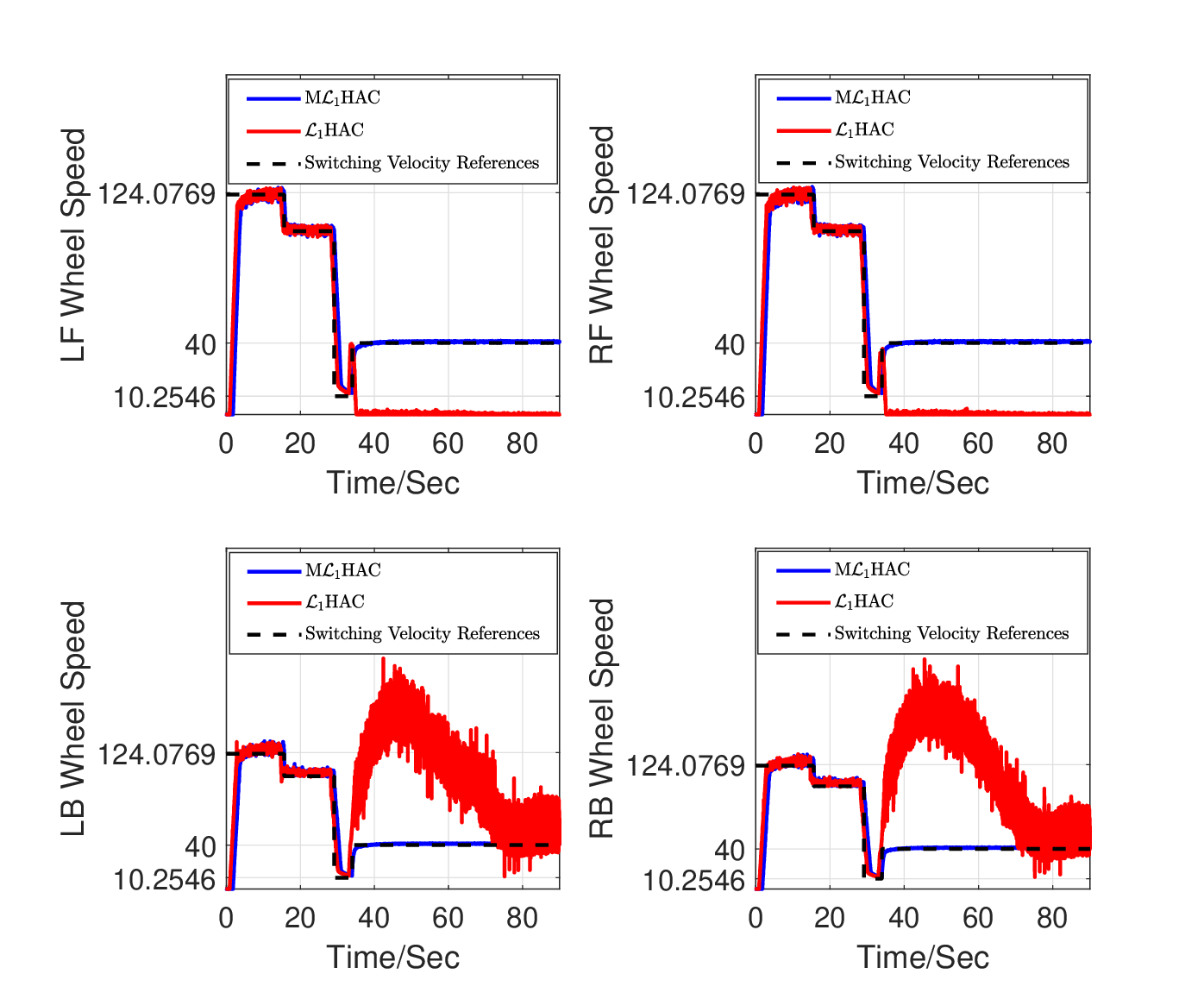}
  \caption{Wheel Angular Velocities}\label{fig:ssa2}
\endminipage\hfill
\minipage{0.50\textwidth}
  \includegraphics[width=\linewidth]{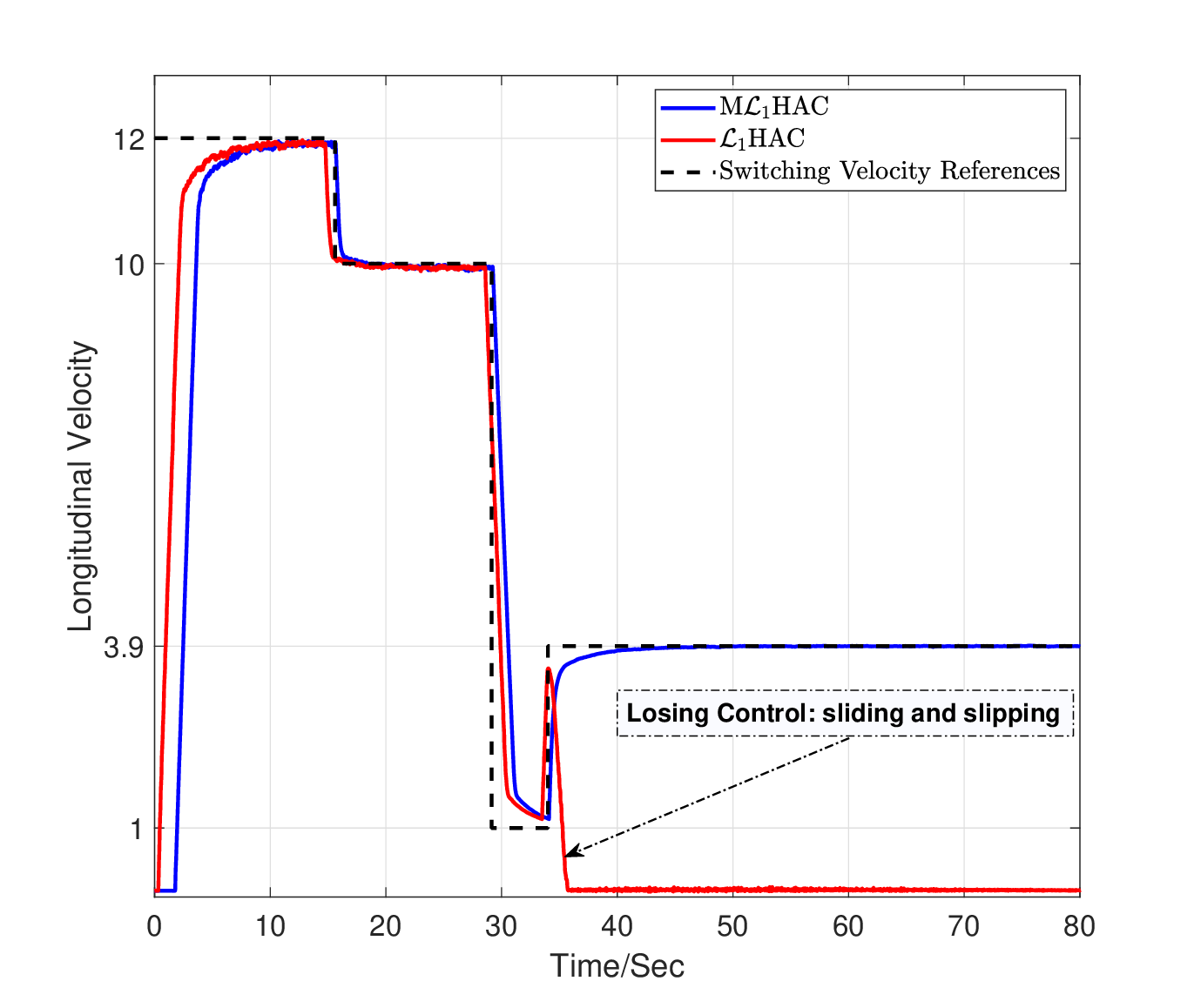}
  \caption{Longitudinal Velocities}\label{fig:ssa3}
\endminipage\hfill
\minipage{0.50\textwidth}%
  \includegraphics[width=\linewidth]{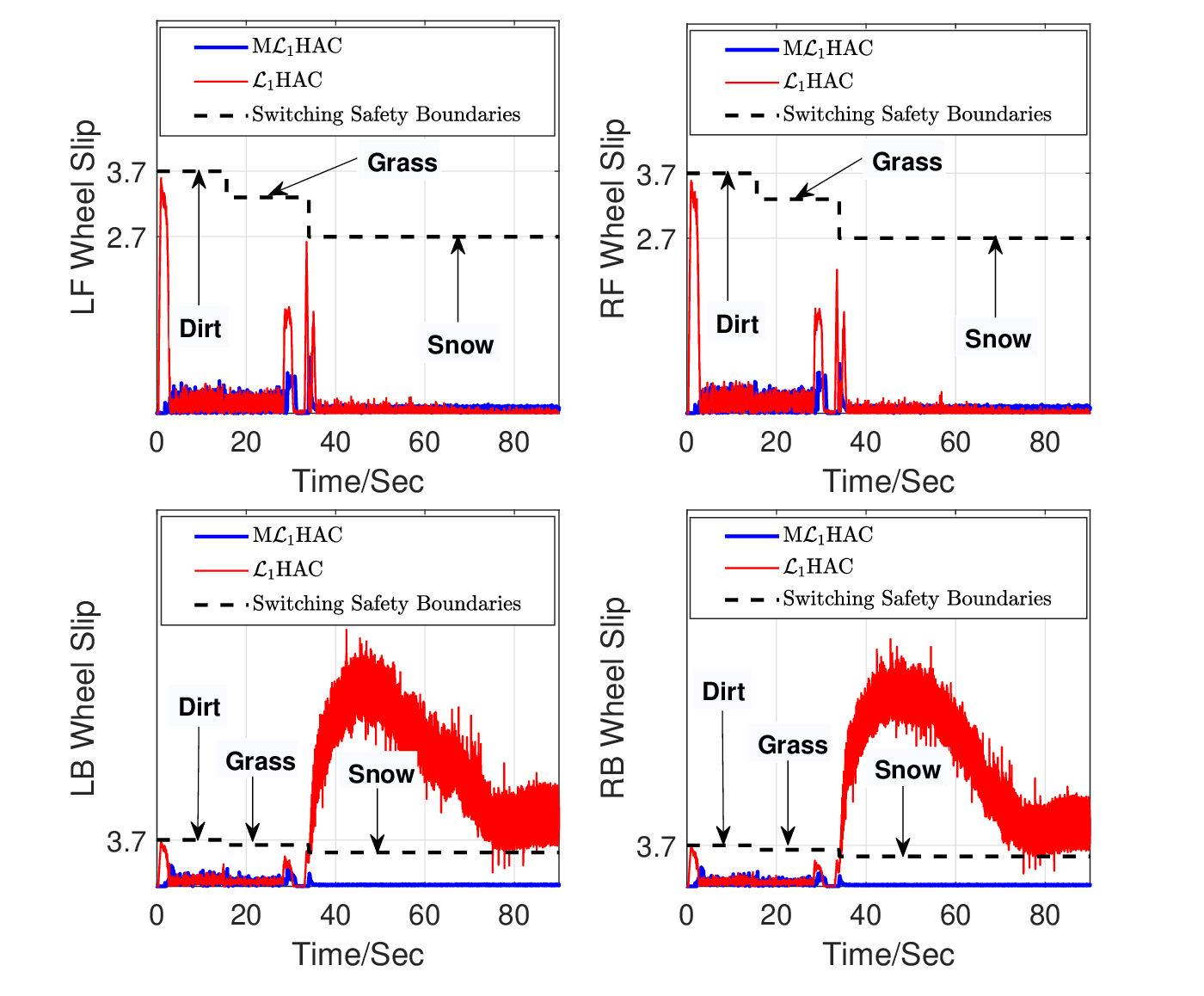}
  \caption{Wheel Slips}\label{fig:ssa4}
\endminipage
\end{figure}

In the experiment, we demonstrate the safe velocity regulation in the dynamic and unforeseen environments via M$\mathcal{L}_{1}$HAC. As shown in Fig. \ref{fig:ssa1}, the vehicle will drive from the dirt and grass areas to the snow area, and the snow area is the unforeseen environment that the vehicle never drove therein before and thus does not have the corresponding sensor data before entering into it.

As Assumption \ref{asm2} states  the environmental perception will accurately  detect the unforeseen snow area in advance. To achieve the safe velocity regulation in the dynamic and unforeseen environments, the safe operation is organized as follows.
\begin{itemize}
  \item The safety boundaries of slip in the dirt, grass and snow areas are set as $\mu_{\mathrm{dirt}}$ = 3.7m/sec, $\mu_{\mathrm{grass}}$ = 3.3m/sec and $\mu_{\mathrm{snow}}$ = 2.7m/sec, respectively. For the state bias, we let $\varepsilon = 2$.
  \item The velocity references of dirt area are set to $[\mathbf{w}^{r}_{\text{dirt}}, \mathbf{v}^{r}_{\text{dirt}}]$ = [123.6rad/sec, 12m/sec].
  \item The velocity references of grass area are initially set to $[\mathbf{w}^{r}_{\text{grass}}, \mathbf{v}^{r}_{\text{grass}}]$ = [103rad/sec, 10m/sec].
  \item The velocity references of grass area is set to $[\mathbf{w}^{r}_{\text{grass}}, \mathbf{v}^{r}_{\text{grass}}]$ = [10.3rad/sec, 1m/sec] 7m ahead of the snow area.
  \item Once the vehicle enters into the snow area, the sensor data in the  first 2 seconds is used to learn the vehicle model, which will guarantee the prescribed levels of accuracy $\phi = 0.9$ and confidence $1 - \delta = 0.7$.
  \item Once the learned model is available, M$\mathcal{L}_{1}$HAC immediate updates the vehicle model with ${A_{\text{learned}}} = \left[ {\begin{array}{*{20}{c}}
  {{\text{0}}{\text{.2444}}}&{ - {\text{0}}{\text{.5151}}} \\
  {{\text{17}}{\text{.0038}}}&{ - {\text{17}}{\text{.0038}}}
\end{array}} \right]$ \& ${B_{\text{learned}}} = \left[ {\begin{array}{*{20}{c}}
  {{\text{0}}{\text{.1}}}&{0.1} \\
  {\text{0}}&0
\end{array}} \right]$.
 \item The controller matrix in the $\mathcal{L}_{1}$ is updated with ${F_{\text{learned}}} = \left[ {\begin{array}{*{20}{c}}
  {{\text{7}}{\text{.9654}}}&{ - {\text{24}}{\text{.5730}}} \\
  {{\text{6}}{\text{.0458}}}&{ - {\text{20}}{\text{.5564}}}
\end{array}} \right]$.
 \item Based on the learned vehicle model, velocity references of snow area are immediately updated with  $[\mathbf{w}^{r}_{\text{snow}}, \mathbf{v}^{r}_{\text{snow}}]$ = [40rad/sec, 3.9m/sec].
\end{itemize}

The trajectories of angular velocities, longitudinal velocities and slip are, respectively, shown in Figs. \ref{fig:ssa2}-\ref{fig:ssa4}, which demonstrate that
\begin{itemize}
  \item The proposed M$\mathcal{L}_{1}$HAC succeeds in safe velocity regulation in the dynamic and unforeseen environments, i.e., the vehicle's angular and longitudinal velocities successfully track the provided switching references, and the four wheel slips are always below the switching safety boundaries.
  \item The $\mathcal{L}_{1}$HAC proposed in \cite{wang2018rsimplex}, i.e., the $\mathcal{L}_{1}$ controller without model learning, fails to maintain safe velocity regulation in the unforeseen snow environment, which is due to the large model mismatch.
\end{itemize}
The demonstration video is available at \textcolor[rgb]{1.00,0.00,1.00}{\url{https://ymao578.github.io/pubs/m1.mp4}}.

\section{Conclusion}
In this paper, we have proposed a novel Simplex architecture for safe velocity regulation of self-driving vehicles through the integration of TCS and ABS. To make the Simplex more reliable in the unprepared or unforeseen environments, finite-time model learning, in conjunction with safe switching control, is incorporated into $\mathcal{L}_{1}$-based verified safe control. The short-term sensor data of vehicle state from a single trajectory is used to adaptively update vehicle model for reliable control actuation computation. Experiments performed in the AutoRally platform demonstrate the effectiveness of the model-learning based $\mathcal{L}_{1}$-Simplex for longitudinal vehicle control systems.

Exploring the model-learning based $\mathcal{L}_{1}$-Simplex in coordinating lateral motion control and longitudinal motion control of self-driving vehicles, as well as the demonstrations in full-size car, constitute our future research directions.

\section*{Acknowledgments}
This work was supported by NSF (award numbers CMMI-1663460, ECCS-1739732 and CPS-1932529).

\bibliographystyle{ACM-Reference-Format}
\bibliography{sample-base}

\end{document}